\documentclass[aps,prl,10pt,twocolumn]{revtex4-2}

\usepackage[T1]{fontenc}
\usepackage{lmodern}
\usepackage{microtype}
\usepackage{amsmath,amssymb,mathtools}

\usepackage{graphicx}
\usepackage{placeins}
\usepackage{xcolor}
\usepackage{tikz}
\usepackage{multirow}
\usetikzlibrary{calc,positioning,arrows.meta}

\usepackage[
  colorlinks=true,
  citecolor=blue!55!black,
  linkcolor=blue!55!black,
  urlcolor=blue!55!black,
  pdftitle={A Combinatorial Origin of Locality},
  pdfauthor={L. Rodina, G. Sellaroli, and J. Valenzuela}
]{hyperref}

\definecolor{cutblue}{RGB}{43,100,176}
\definecolor{cutred}{RGB}{190,55,55}
\definecolor{cutgray}{RGB}{82,91,103}
\definecolor{cutgreen}{RGB}{38,130,82}
\definecolor{cutpurple}{RGB}{124,75,154}
\definecolor{cutorange}{RGB}{215,128,42}
\tikzset{
  casepoly/.style={
    line width=.55pt,
    draw=cutgray
  },
  casesupp/.style={
    line width=1.05pt,
    draw=cutblue
  },
  caseleaf/.style={
    line width=1.25pt,
    draw=cutred
  },
  caseshort/.style={
    line width=1.1pt,
    draw=cutpurple
  },
  casescar/.style={
    line width=.95pt,
    draw=cutorange,
    dashed
  },
  casecut/.style={
    line width=.85pt,
    draw=cutred,
    dashed
  },
  caseflow/.style={
    -{Latex[length=1.5mm]},
    line width=.65pt,
    draw=cutgray
  },
  casecenter/.style={
    circle,
    fill=cutgreen,
    draw=cutgreen,
    inner sep=1.35pt
  },
  casepoint/.style={
    circle,
    draw=cutgray,
    fill=white,
    inner sep=.85pt
  },
  alginput/.style={
    draw=cutgray,
    fill=cutgray!5,
    rounded corners=1.5pt,
    line width=.55pt,
    align=center,
    inner xsep=3pt,
    inner ysep=2.5pt,
    font=\scriptsize
  },
  algtest/.style={
    draw=cutorange!90!black,
    fill=cutorange!7,
    rounded corners=1.5pt,
    line width=.65pt,
    align=center,
    inner xsep=3pt,
    inner ysep=2.5pt,
    font=\scriptsize
  },
  algop/.style={
    draw=cutblue,
    fill=cutblue!5,
    rounded corners=1.5pt,
    line width=.55pt,
    align=center,
    inner xsep=3pt,
    inner ysep=2.5pt,
    font=\scriptsize
  },
  algreturn/.style={
    draw=cutgreen!85!black,
    fill=cutgreen!7,
    rounded corners=1.5pt,
    line width=.7pt,
    align=center,
    inner xsep=3pt,
    inner ysep=2.5pt,
    font=\scriptsize
  },
  algarrow/.style={
    -{Latex[length=1.35mm,width=1.1mm]},
    draw=cutgray,
    line width=.55pt
  },
  algbranch/.style={
    font=\scriptsize\bfseries,
    text=cutgray,
    fill=white,
    inner sep=1pt
  }
}

\newcommand{\Trphi}{\operatorname{Tr}(\phi^3)}

\newcommand{\cB}{\mathcal B}

\begin{document}

\title{A Combinatorial Origin of Locality}

\author{Laurentiu Rodina$^1$}
\email{laurentiu.rodina@gmail.com}
\author{Giuseppe Sellaroli$^2$}
\email{gsellaro@uwaterloo.ca}
\author{Yan Ni (Jonathan Valenzuela)$^3$}
\email{jonathanalbert10@mails.tsinghua.edu.cn}
% Add institutional affiliations before submission.\\

\affiliation{$\mbox{}^{1}$Beijing Institute of Mathematical Sciences and Applications (BIMSA), Beijing, 101408, China}
\affiliation{$\mbox{}^{2}$University of Waterloo, Waterloo, Ontario N2L 3G1, Canada}
\affiliation{$\mbox{}^{3}$Tsinghua University, Beijing, 100084, China}

\begin{abstract}
Locality and unitarity are fundamental principles of quantum field theory.
For tree-level scattering amplitudes, locality determines which propagator
poles may occur together, while unitarity fixes their residues through
factorization.  Previous uniqueness theorems have repeatedly shown that
unitarity can emerge from locality combined with other physical principles.
In this Letter we give the first complete all-multiplicity proof for the
emergence of locality in this setting.  We focus on Tr($\phi^3$) theory and phrase locality as a problem in geometric
combinatorics. Planar propagators are
chords of a polygon, and a denominator is local exactly when its chords
triangulate the polygon, and non-local otherwise.  We show that hidden zeros---regular kinematic
conditions on which the amplitude vanishes---enforce pole compatibility and therefore imply a local singularity structure. The proof rests on two ingredients:
a graphical \emph{star test} determines when a chosen zero eliminates a pole
product, and a \emph{smoothing} procedure finds such a zero.
Thus, once the planar pole alphabet and denominator bound are specified,
locality and unitarity emerge together from a single on-shell principle:
hidden zeros.  Equivalently, our proof provides a novel characterization of a familiar concept:
polygon triangulations are precisely the chord
configurations that evade every  star test.
\end{abstract}

\maketitle

\section{Introduction}

The Lagrangian formulation builds quantum field theory from interactions
localized in spacetime and a unitary time evolution.  At tree level, locality
means that the propagators appearing together in a pole product must belong
to a single Feynman graph, while unitarity requires factorization on those
poles.  Modern on-shell and geometric formulations raise a sharper question: must
the local graph structure be assumed, or can locality and unitarity emerge
from a different starting point
\cite{Amplituhedron,PositiveGeometries,ABHY,GaugeLocality}?  Deriving both locality
and unitarity would point toward formulations of quantum field theory based
on principles different from those of a local Lagrangian.

Many uniqueness results have already shown that unitarity can emerge.
Starting from a local pole ansatz, gauge invariance, soft behavior, or
ultraviolet scaling can uniquely fix the amplitude and recover its correct
factorization
\cite{GaugeLocality,AdlerUniqueness,SoftUniqueness,
RodinaBCFW,CarrascoRodinaUV,RodinaUV}.  Locality remained an input in the
all-multiplicity proofs, although finite-multiplicity evidence suggested that
it too could emerge \cite{GaugeLocality,HiddenZeros,BackusRodina} (see also related developments~\cite{
NLSMRecursionOld,NLSMTree,OnShellRecursionEFT,
EFTPeriodicTable,CheungKampfNovotnyTrnka}).

In this Letter we give the first complete all-multiplicity proof that
locality can itself emerge from other on-shell properties.  To isolate the heart of
the problem---the compatibility of propagator poles---we work with
color-ordered \(\Trphi\) $n$-point tree amplitudes.  Despite its simplicity, this theory
retains the full combinatorics of planar poles, while its constant numerators
remove the additional structures that complicate the nonlinear sigma model (NLSM) and Yang--Mills
theory.  We bootstrap it from the minimal data needed for a finite ansatz: the
planar pole alphabet, arbitrary constant coefficients, and a denominator
bound of at most \(n-3\) factors.  We assume neither the exact pole count,
simple poles, nor compatibility, and impose no factorization or residue data.

What remains is to find a bootstrap principle capable of determining both
which poles may coexist and their relative coefficients.   Hidden
zeros \cite{HiddenZeros} (see also
\cite{HiddenZerosDoubleCopy,ScaffoldedGravityZeros,UniversalZeros,
ZhouDiagramZeros,FengBCFWZeros,SmoothSplittingZeros,NLSMRecursion,
HiddenZerosMassive,HiddenZerosHigherDerivative,CosmologicalHiddenZeros,CosmologicalWavefunctionZeros})---regular kinematic conditions on which the amplitude vanishes without
requiring a physical propagator to go on shell---provide precisely such a
principle.  Although defined without reference to poles or factorization, hidden zeros
fix amplitudes within a local ansatz and constrain nonlocal terms at finite
multiplicity \cite{HiddenZeros,RodinaUV,BackusRodina}.  Here we prove this second statement
at all multiplicities and uncover the concrete mechanism by which locality
emerges.

For a fixed cyclic ordering, each planar channel
\[
 X_{ij}=(p_i+p_{i+1}+\cdots+p_{j-1})^2
\]
is represented by a chord joining vertices \(i\) and \(j\) of an
\(n\)-gon.  The \(n-3\) propagators of a planar cubic tree triangulate the
polygon, and every triangulation is dual to one such tree.  A pole product is
therefore local precisely when its chords form a triangulation.  Motivated by
surfaceology
\cite{AllLoopCounting,AllLoopMultiplicity,HiddenZeros}, we recast the emergence of locality
as a problem in geometric combinatorics.  This simplest setting exposes the
fundamental mechanism cleanly and provides a starting point for the NLSM and
Yang--Mills theory, where nontrivial numerators obscure the same
pole-compatibility problem.

We prove that every nontriangulating chord configuration is directly
incompatible with at least one hidden zero and, more strongly, give a
constructive procedure for finding that zero.  A graphical \emph{star test}
determines whether a proposed zero eliminates a given pole product, while a
recursive \emph{smoothing} of the chord configuration constructs a zero that
passes the test. 

The zeros therefore do more than fix the coefficients of local diagrams:
they select the diagrams themselves and enforce both the correct pole count
and simple poles.  This strengthens the conjecture of
Ref.~\cite{HiddenZeros}, which  assumed exactly \(n-3\) distinct poles per term.  Our theorem allows
terms with fewer poles and repeated poles and eliminates every such term
constructively.  Together with the earlier uniqueness theorem \cite{RodinaUV}, this fixes the local,
factorizing amplitude up to normalization.  Thus, assuming the pole alphabet, locality and unitarity both
emerge from hidden zeros.

Read in the opposite direction, the theorem also gives a new characterization
of a familiar combinatorial object. Among configurations of at most \(n-3\)
chords, polygon triangulations are precisely those that evade every interval
star test.

\section{A five-point observation}

We work in massless \(\operatorname{Tr}(\phi^3)\) theory, a matrix-valued
scalar theory with a cubic single-trace interaction.  Fixing the trace order
\((1,2,\ldots,n)\) selects a color-ordered \(n\)-point tree amplitude whose
Feynman diagrams are planar.  Their numerators are constant, so this theory
isolates the question of which propagator poles may occur together.

For outgoing momenta satisfying \(\sum_i p_i=0\), $p_i^2=0$, the possible planar
channels are
\begin{equation}
 X_{ij}=(p_i+p_{i+1}+\cdots+p_{j-1})^2,
 \qquad X_{i,i}=X_{i,i+1}=0,
 \label{eq:planar-variable}
\end{equation}
with labels understood cyclically and complementary channels identified.

At five points there are five channels $\{
 X_{13}, X_{24}, X_{35},X_{14}, X_{25}\}$.
Allowing at most two denominator factors, the most general
constant-coefficient ansatz can be written cyclically as
\begin{align}
\nonumber \cB_5=&A_0+\sum_{i=1}^{5}
 \left(
 \frac{b_i}{X_{i,i+2}}
 +\frac{r_i}{X_{i,i+2}^{\,2}}+\frac{\alpha_i}{X_{i,i+2}X_{i+1,i+3}}\right.\\
 &\left.+\frac{c_i}{X_{i,i+2}X_{i+2,i+4}}\right).
 \label{eq:five-ansatz}
\end{align}
The \(c_i\)-terms are the five pole products appearing in planar cubic
trees.  The remaining terms are the non-local ones. They have too few poles, repeated poles, or pairs of
poles that cannot occur together in one such tree.

Now impose the hidden zero (the general zero locus is given in the next section)
\begin{equation}
 H_{45}:\qquad
 X_{35}=-X_{14},
 \qquad
 X_{25}=X_{24}-X_{14}.
 \label{eq:H45}
\end{equation}
This is a regular kinematic locus: \(X_{13},X_{24}\), and \(X_{14}\) remain
generic, and no propagator is required to go on shell.  Substituting Eq.~\eqref{eq:H45} into Eq.~\eqref{eq:five-ansatz} and
requiring the result to vanish identically gives, among other conditions,
\(\alpha_1=0\) and \(c_1=c_4\).  Thus the zero directly eliminates some
nonlocal terms while relating the coefficients of the remaining terms.

Imposing its five cyclic
images gives
\begin{equation}
 A_0=b_i=r_i=\alpha_i=0,
 \qquad
 c_1=c_2=\cdots=c_5,
 \label{eq:five-solution}
\end{equation}
and therefore
\begin{equation}
 \cB_5
 =
 c\sum_{i=1}^{5}
 \frac{1}{X_{i,i+2}X_{i+2,i+4}},
 \label{eq:five-amplitude}
\end{equation}
which is the planar five-point amplitude up to an overall normalization.

This example separates the two effects of hidden zeros. They eliminate
forbidden denominators and relate the coefficients of the surviving
triangulations.  It also poses the all-multiplicity problem:
\begin{center}
 \emph{Given a pole product that belongs to no planar cubic tree, which
 hidden zero eliminates it?}
\end{center}
At five points this can be answered by direct substitution.  To solve the
problem for arbitrary \(n\), we now reformulate it geometrically.

\section{The chord formulation of locality}

Place the labels \(1,\ldots,n\) cyclically on the boundary of a disk and
represent each nontrivial channel \(X_{ij}\) by the chord joining vertices
\(i\) and \(j\).  This is the tree-level disk of surfaceology; the same chord
combinatorics underlies the associahedron and dihedral coordinates
\cite{Stasheff,ABHY,BrownM0n}.

\begin{figure}[t]
\centering
\begin{tikzpicture}[
  font=\scriptsize,
  poly/.style={draw=cutgray,line width=.55pt},
  localchord/.style={draw=cutblue,line width=1.15pt},
  nonlocalchord/.style={draw=cutred,line width=1.15pt},
  dualedge/.style={
    draw=cutgreen!85!black,
    line width=.8pt,
    densely dashed,
    preaction={draw=white,line width=2.15pt}
  },
  dualvertex/.style={
    circle,
    fill=cutgreen!85!black,
    draw=white,
    line width=.35pt,
    inner sep=1.05pt
  }
]

\begin{scope}[shift={(-1.72,0)}]
  \foreach \ang/\lab in {90/1,18/2,-54/3,-126/4,162/5}{
    \coordinate (p\lab) at (\ang:.78);
    \node at ($(p\lab)+(\ang:.115)$) {\lab};
  }
  \coordinate (t123) at (barycentric cs:p1=1,p2=1,p3=1);
  \coordinate (t134) at (barycentric cs:p1=1,p3=1,p4=1);
  \coordinate (t145) at (barycentric cs:p1=1,p4=1,p5=1);
  \coordinate (m12) at ($(p1)!.5!(p2)$);
  \coordinate (m23) at ($(p2)!.5!(p3)$);
  \coordinate (m34) at ($(p3)!.5!(p4)$);
  \coordinate (m45) at ($(p4)!.5!(p5)$);
  \coordinate (m51) at ($(p5)!.5!(p1)$);

  \draw[poly] (p1)--(p2)--(p3)--(p4)--(p5)--cycle;
  \draw[localchord] (p1)--(p3) (p1)--(p4);
  \draw[dualedge]
    (m12)--(t123)--(m23)
    (t123)--(t134)--(t145)
    (t134)--(m34)
    (t145)--(m45)
    (t145)--(m51);
  \node[dualvertex] at (t123) {};
  \node[dualvertex] at (t134) {};
  \node[dualvertex] at (t145) {};

  \node[font=\scriptsize\bfseries,text=cutgreen!70!black]
    at (0,1.08) {(a) local};
  \node at (0,-1.06) {$\displaystyle \frac{1}{X_{13}X_{14}}$};
\end{scope}

\begin{scope}[shift={(1.72,0)}]
  \foreach \ang/\lab in {90/1,18/2,-54/3,-126/4,162/5}{
    \coordinate (q\lab) at (\ang:.78);
    \node at ($(q\lab)+(\ang:.115)$) {\lab};
  }
  \draw[poly] (q1)--(q2)--(q3)--(q4)--(q5)--cycle;
  \draw[nonlocalchord] (q1)--(q3) (q2)--(q4);

  \node[font=\scriptsize\bfseries,text=cutred]
    at (0,1.08) {(b) nonlocal};
  \node at (0,-1.06) {$\displaystyle \frac{1}{X_{13}X_{24}}$};
  \node[anchor=west,align=left,font=\tiny,text=cutred]
    at (1.00,.02) {no planar\\dual tree};
\end{scope}
\end{tikzpicture}
\caption{\label{fig:local-nonlocal}
Five-point locality.  Solid chords denote propagators.  In (a), the green
dashed graph drawn over the triangulation is its dual planar cubic tree.  In
(b), the crossing chords admit no planar dual.}
\end{figure}
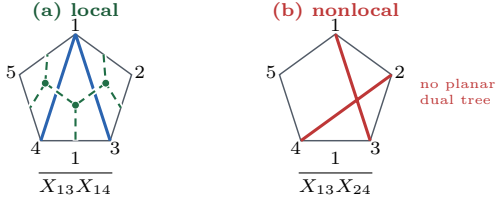

A planar cubic tree contains \(n-3\) internal propagators.  Its corresponding
chords are distinct, do not cross, and triangulate the polygon.  Conversely,
the graph dual to every triangulation is a planar cubic tree.  Thus, within
the cubic-tree ansatz, locality has a purely geometric formulation: a pole
product is local precisely when its chords form a triangulation.
Figure~\ref{fig:local-nonlocal} illustrates this correspondence at five
points.

At general multiplicity, let \(M\) be a multiset of chords and consider
\begin{align}
 \cB_n&=
 \sum_{\substack{M\text{ multiset of chords}\\ |M|\leq n-3}}
 \frac{a_M}{X_M},
 \notag\\[-2pt]
 X_M&=\prod_{ij\in M}X_{ij},
 \qquad
 S=\operatorname{supp}M .
 \label{eq:ansatz}
\end{align}
Using a multiset allows repeated poles, while the bound permits terms with
fewer than \(n-3\) factors.  We call \(M\) \emph{local} when it consists of
exactly \(n-3\) distinct chords forming a triangulation, and
\emph{nonlocal} otherwise. We call the chords in \(S\) selected.

The hidden zeros are labelled by consecutive boundary intervals
\begin{equation}
 T=\{w,w+1,\ldots,w+k\},
 \qquad 2\leq |T|\leq n-3 .
 \label{eq:zero-interval}
\end{equation}
We call \(T\) the \emph{zero interval}.  Writing \(v=w+k+1\), the
corresponding locus is
\begin{equation}
 \begin{gathered}
 H_T:\quad
 X_{w+i,j}
 =
 X_{w,j}-X_{w,v}+X_{w+i,v},
 \\[-2pt]
 1\leq i\leq k,\qquad j\in T^c .
 \end{gathered}
 \label{eq:zero}
\end{equation}
At a generic point of \(H_T\), no propagator channel is required to vanish.
We impose
\begin{equation}
 \left.\cB_n\right|_{H_T}=0
 \qquad\text{for every zero interval }T .
 \label{eq:all-zero-condition}
\end{equation}

The problem is now entirely geometric. Given a chord multiset that does not
triangulate the polygon, find a boundary interval whose zero isolates its
coefficient.  For \(n\geq5\), we prove that such an interval always exists.
Consequently, imposing all hidden zeros eliminates every nonlocal term and
leaves only the denominators of planar cubic trees.  More strongly, for each
nonlocal pole product \(M\), we construct a zero interval \(T\) that isolates
\(a_M\) and forces it to vanish. A separate question is to ask for the minimal number of zeros that accomplishes this fact. We leave this question
for future work.

\section{Proof strategy: why crossings are a red herring}

A crossing pair makes nonlocality immediately visible, but it turns out a zero
cannot detect the crossing of chords itself.  Instead, it detects a gap. A region of
the polygon that has not been correctly triangulated.  At the full cubic pole
count, placing a chord in an incompatible position necessarily leaves a
triangulating chord missing elsewhere. A crossing then is useful only insofar  it signals a chord is missing from its rightful position. So in principle, we need only a geometric mechanism by which a zero detects a non-triangulated region.

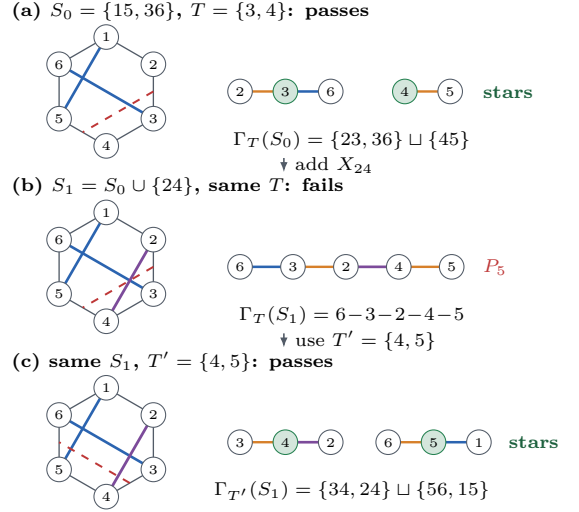
\begin{figure}[t]
\centering
\begin{tikzpicture}[
  x=1cm,y=1cm,
  font=\scriptsize,
  pvertex/.style={circle,draw=cutgray,fill=white,
    minimum size=3.25mm,inner sep=0pt,font=\tiny},
  gvertex/.style={circle,draw=cutgray,fill=white,
    minimum size=3.25mm,inner sep=0pt,font=\tiny},
  gcenter/.style={circle,draw=cutgreen,fill=cutgreen!20,
    minimum size=3.25mm,inner sep=0pt,font=\tiny},
  poly/.style={draw=cutgray,line width=.5pt},
  oldchord/.style={draw=cutblue,line width=1pt},
  newchord/.style={draw=cutpurple,line width=1.05pt},
  cut/.style={draw=cutred,dashed,line width=.75pt},
  gselected/.style={draw=cutblue,line width=.95pt},
  gnew/.style={draw=cutpurple,line width=1pt},
  gboundary/.style={draw=cutorange,line width=.9pt},
  pollflow/.style={-{Latex[length=1.35mm]},draw=cutgray,line width=.6pt}
]

\node[anchor=west,font=\scriptsize\bfseries] at (-3.72,7.08)
  {(a) \(S_0=\{15,36\}\), \(T=\{3,4\}\): passes};
\begin{scope}[shift={(-2.35,6.02)}]
  \foreach \i/\ang in {1/90,2/30,3/-30,4/-90,5/-150,6/150}
    \coordinate (A\i) at (\ang:.72);
  \draw[poly] (A1)--(A2)--(A3)--(A4)--(A5)--(A6)--cycle;
  \draw[oldchord] (A1)--(A5);
  \draw[oldchord] (A3)--(A6);
  \draw[cut] ($(A2)!.5!(A3)$)--($(A4)!.5!(A5)$);
  \foreach \i in {1,...,6} \node[pvertex] at (A\i) {\i};
\end{scope}
\coordinate (Ag2) at (-.58,6.02);
\coordinate (Ag3) at ( .02,6.02);
\coordinate (Ag6) at ( .62,6.02);
\coordinate (Ag4) at (1.60,6.02);
\coordinate (Ag5) at (2.20,6.02);
\draw[gboundary] (Ag2)--(Ag3);
\draw[gselected] (Ag3)--(Ag6);
\draw[gboundary] (Ag4)--(Ag5);
\node[gvertex] at (Ag2) {2};
\node[gcenter] at (Ag3) {3};
\node[gvertex] at (Ag6) {6};
\node[gcenter] at (Ag4) {4};
\node[gvertex] at (Ag5) {5};
\node[anchor=west,text=cutgreen!75!black,font=\scriptsize\bfseries]
  at (2.52,6.02) {stars};
\node at (.90,5.38)
  {\(\Gamma_T(S_0)=\{23,36\}\sqcup\{45\}\)};

\draw[pollflow] (0,5.14)--(0,4.95)
  node[midway,right=3pt,fill=white,inner sep=1pt] {add \(X_{24}\)};

\node[anchor=west,font=\scriptsize\bfseries] at (-3.72,4.76)
  {(b) \(S_1=S_0\cup\{24\}\), same \(T\): fails};
\begin{scope}[shift={(-2.35,3.70)}]
  \foreach \i/\ang in {1/90,2/30,3/-30,4/-90,5/-150,6/150}
    \coordinate (B\i) at (\ang:.72);
  \draw[poly] (B1)--(B2)--(B3)--(B4)--(B5)--(B6)--cycle;
  \draw[oldchord] (B1)--(B5);
  \draw[oldchord] (B3)--(B6);
  \draw[newchord] (B2)--(B4);
  \draw[cut] ($(B2)!.5!(B3)$)--($(B4)!.5!(B5)$);
  \foreach \i in {1,...,6} \node[pvertex] at (B\i) {\i};
\end{scope}
\coordinate (Bg6) at (-.58,3.70);
\coordinate (Bg3) at ( .12,3.70);
\coordinate (Bg2) at ( .82,3.70);
\coordinate (Bg4) at (1.52,3.70);
\coordinate (Bg5) at (2.22,3.70);
\draw[gselected] (Bg6)--(Bg3);
\draw[gboundary] (Bg3)--(Bg2);
\draw[gnew] (Bg2)--(Bg4);
\draw[gboundary] (Bg4)--(Bg5);
\node[gvertex] at (Bg6) {6};
\node[gvertex] at (Bg3) {3};
\node[gvertex] at (Bg2) {2};
\node[gvertex] at (Bg4) {4};
\node[gvertex] at (Bg5) {5};
\node[anchor=west,text=cutred,font=\scriptsize\bfseries]
  at (2.52,3.70) {$P_5$};
\node at (.90,3.06) {\(\Gamma_T(S_1)=6\!-\!3\!-\!2\!-\!4\!-\!5\)};

\draw[pollflow] (0,2.82)--(0,2.63)
  node[midway,right=3pt,fill=white,inner sep=1pt] {use \(T'=\{4,5\}\)};

\node[anchor=west,font=\scriptsize\bfseries] at (-3.72,2.44)
  {(c) same \(S_1\), \(T'=\{4,5\}\): passes};
\begin{scope}[shift={(-2.35,1.38)}]
  \foreach \i/\ang in {1/90,2/30,3/-30,4/-90,5/-150,6/150}
    \coordinate (C\i) at (\ang:.72);
  \draw[poly] (C1)--(C2)--(C3)--(C4)--(C5)--(C6)--cycle;
  \draw[oldchord] (C1)--(C5);
  \draw[oldchord] (C3)--(C6);
  \draw[newchord] (C2)--(C4);
  \draw[cut] ($(C3)!.5!(C4)$)--($(C5)!.5!(C6)$);
  \foreach \i in {1,...,6} \node[pvertex] at (C\i) {\i};
\end{scope}
\coordinate (Cg3) at (-.58,1.38);
\coordinate (Cg4) at ( .02,1.38);
\coordinate (Cg2) at ( .62,1.38);
\coordinate (Cg6) at (1.38,1.38);
\coordinate (Cg5) at (1.98,1.38);
\coordinate (Cg1) at (2.58,1.38);
\draw[gboundary] (Cg3)--(Cg4);
\draw[gnew] (Cg4)--(Cg2);
\draw[gboundary] (Cg6)--(Cg5);
\draw[gselected] (Cg5)--(Cg1);
\node[gvertex] at (Cg3) {3};
\node[gcenter] at (Cg4) {4};
\node[gvertex] at (Cg2) {2};
\node[gvertex] at (Cg6) {6};
\node[gcenter] at (Cg5) {5};
\node[gvertex] at (Cg1) {1};
\node[anchor=west,text=cutgreen!75!black,font=\scriptsize\bfseries]
  at (2.86,1.38) {stars};
\node at (.90,.74)
  {\(\Gamma_{T'}(S_1)=\{34,24\}\sqcup\{56,15\}\)};

\end{tikzpicture}
\caption{A misplaced chord can hide a gap from one zero.  The cut \(T\)
passes for \(S_0\), but adding the crossing chord \(X_{24}\) joins its two
stars into a $P_5$.  The enlarged support \(S_1\) is detected by the
different cut \(T'\).  Red dashes mark the cut; blue and purple links are
selected chords, and orange links are its boundary links.}
\label{fig:polluted-cut}
\end{figure}

What makes this even more difficult is that other
misplaced chords can obscure this gap from a particular zero, making the
region appear triangulated even though the full configuration is not, as illustrated in Fig.~\ref{fig:polluted-cut}.  The central
task is therefore to prove that \(n-3\) chords are not enough to hide every
gap from every interval zero. At least one zero must still expose the
nonlocality.  Consequently, the successful zero depends on the complete
chord configuration and cannot, in general, be found by looking only at an
obvious crossing pair.  This is what makes the proof more non-trivial than one might naively expect given its simple statement.

\begin{figure}[t]
\centering
\begin{tikzpicture}[
  x=1cm,y=1cm,
  font=\scriptsize,
  ovpoint/.style={circle,fill=cutgray,draw=cutgray,inner sep=.85pt},
  ovhead/.style={font=\scriptsize,text=cutgray,align=center},
  ovcase/.style={font=\scriptsize\bfseries}
]

% F0
\begin{scope}[shift={(-3.0,0)}]
  \coordinate (a) at (102:.55);
  \coordinate (p) at (42:.55);
  \coordinate (b) at (-18:.55);
  \draw[casepoly] (0,0) circle (.55);
  \foreach \q in {a,p,b} \node[ovpoint] at (\q) {};
  \node at ($(a)+(-.10,.14)$) {$a$};
  \node at ($(p)+(.13,.11)$) {$p$};
  \node at ($(b)+(.13,-.01)$) {$b$};
  \node[ovhead] at (0,.88) {free};
  \node[ovcase] at (0,-.76) {F0};
  \node[ovhead] at (0,-1.02) {short absent};
\end{scope}

% F1
\begin{scope}[shift={(-1.0,0)}]
  \coordinate (a) at (102:.55);
  \coordinate (p) at (42:.55);
  \coordinate (b) at (-18:.55);
  \draw[casepoly] (0,0) circle (.55);
  \draw[caseshort] (a)--(b);
  \foreach \q in {a,p,b} \node[ovpoint] at (\q) {};
  \node at ($(a)+(-.10,.14)$) {$a$};
  \node at ($(p)+(.13,.11)$) {$p$};
  \node at ($(b)+(.13,-.01)$) {$b$};
  \node[ovhead] at (0,.88) {free};
  \node[ovcase] at (0,-.76) {F1};
  \node[ovhead] at (0,-1.02) {short present};
\end{scope}

% L0
\begin{scope}[shift={(1.0,0)}]
  \coordinate (a) at (102:.55);
  \coordinate (p) at (42:.55);
  \coordinate (b) at (-18:.55);
  \coordinate (u) at (220:.55);
  \draw[casepoly] (0,0) circle (.55);
  \draw[caseleaf] (p)--(u);
  \foreach \q in {a,p,b,u} \node[ovpoint] at (\q) {};
  \node at ($(a)+(-.10,.14)$) {$a$};
  \node at ($(p)+(.13,.11)$) {$p$};
  \node at ($(b)+(.13,-.01)$) {$b$};
  \node at ($(u)+(-.12,-.12)$) {$u$};
  \node[ovhead] at (0,.88) {leaf};
  \node[ovcase] at (0,-.76) {L0};
  \node[ovhead] at (0,-1.02) {short absent};
\end{scope}

% L1
\begin{scope}[shift={(3.0,0)}]
  \coordinate (a) at (102:.55);
  \coordinate (p) at (42:.55);
  \coordinate (b) at (-18:.55);
  \coordinate (u) at (220:.55);
  \draw[casepoly] (0,0) circle (.55);
  \draw[caseshort] (a)--(b);
  \draw[caseleaf] (p)--(u);
  \foreach \q in {a,p,b,u} \node[ovpoint] at (\q) {};
  \node at ($(a)+(-.10,.14)$) {$a$};
  \node at ($(p)+(.13,.11)$) {$p$};
  \node at ($(b)+(.13,-.01)$) {$b$};
  \node at ($(u)+(-.12,-.12)$) {$u$};
  \node[ovhead] at (0,.88) {leaf};
  \node[ovcase] at (0,-.76) {L1};
  \node[ovhead] at (0,-1.02) {short present};
\end{scope}

\end{tikzpicture}
\caption{The four local configurations.  The vertices \(a,p,b\) are
consecutive.  In F0 and F1, \(p\) is free.  In L0 and L1, the red chord
\(X_{pu}\) is the unique selected chord ending at \(p\).  Purple denotes the
short chord \(X_{ab}\); all other chords are suppressed.}
\label{fig:four-local-cases}
\end{figure}
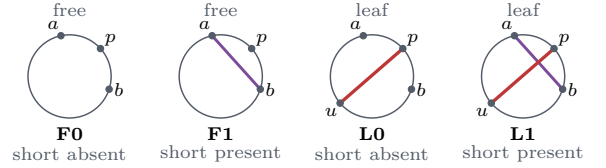

We solve this problem in two steps.  A graphical \emph{star test} first
decides whether a proposed hidden zero isolates the chosen denominator.  We
then find a zero passing this test by recursively removing one simple vertex (free vertex or leaf vertex).
Only the four local configurations in Fig.~\ref{fig:four-local-cases} can
occur.  The main text gives the logic of the recursion; the explicit cuts,
lifts, and boundary cases are collected in Appendix~2.

\section{Zeros and stars as graphical detectors}

Given a pole product and a candidate hidden zero, we need to determine
whether the zero isolates---and therefore eliminates---that term, without
substituting it into the full ansatz.  To make the mechanism transparent,
we first assume that every term contains exactly \(n-3\) distinct, simple
poles.   In generic kinematics, \(a_M\) is extracted by taking the
simultaneous residue in all poles belonging to \(M\):
\begin{equation}
 \underset{M}{\operatorname{Res}}\,\cB_n=a_M.
 \label{eq:generic-residue}
\end{equation}
If the same residue remains isolated after restriction to \(H_T\), then the
zero condition immediately gives
\begin{equation}
 0=\underset{M}{\operatorname{Res}}
 \left(\left.\cB_n\right|_{H_T}\right)=a_M.
 \label{eq:clean-residue}
\end{equation}

The restriction can, however, force an additional channel
\(X_d\notin M\) to vanish together with the poles in \(M\).  A competing
term \(M'\) containing \(X_d\) may then acquire the same multiple pole, so
that, schematically,
\begin{equation}
 \underset{M}{\operatorname{Res}}
 \left(\left.\cB_n\right|_{H_T}\right)=a_M+a_{M'}.
 \label{eq:polluted-residue}
\end{equation}
The zero now relates two coefficients instead of eliminating \(a_M\).  We
call this \emph{pollution}.  The purpose of the star test is to recognize,
directly from the chords, when the pollution does not occur.

\begin{figure}
    \centering
    \includegraphics[width=0.75\linewidth]{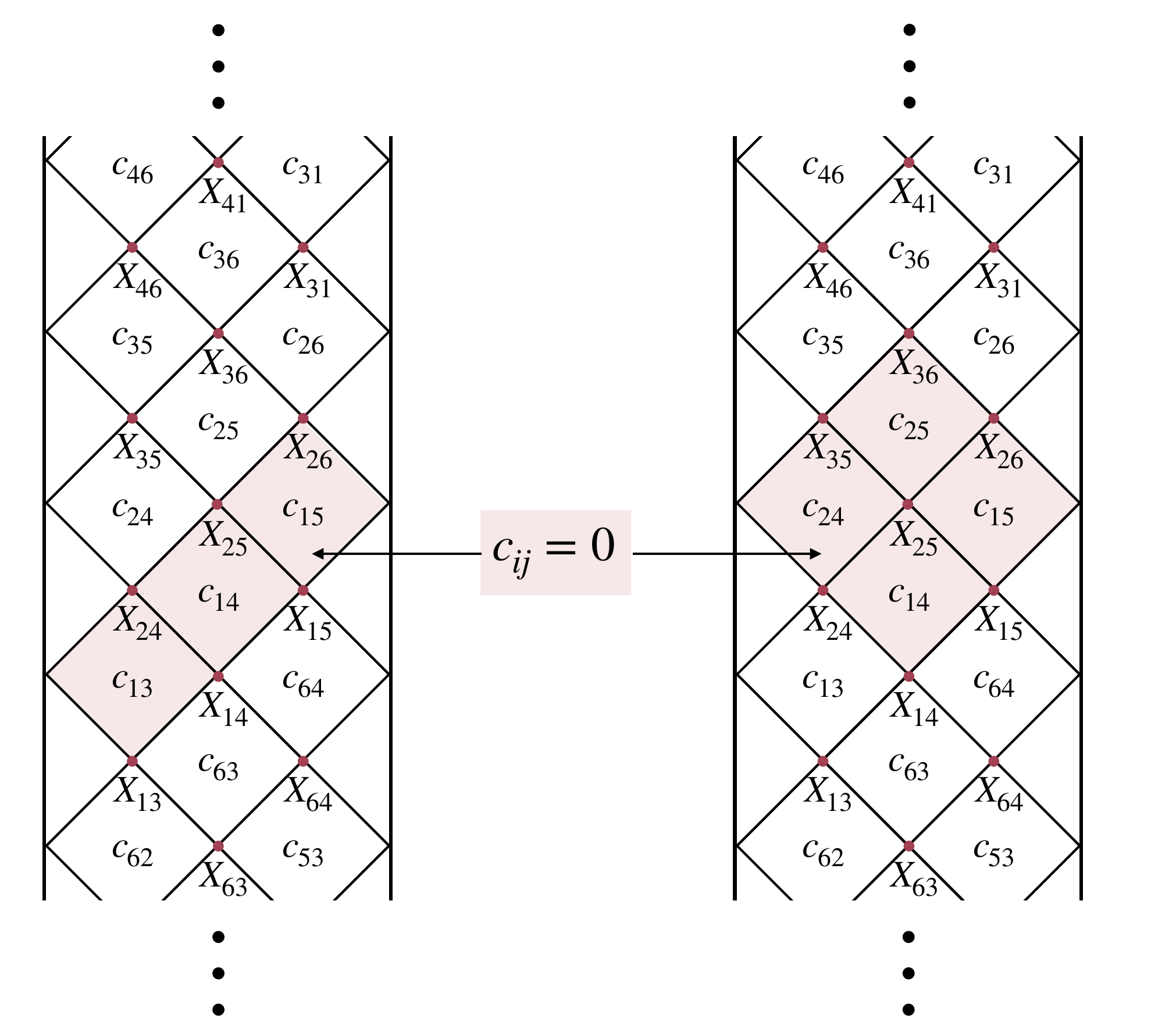}
    \caption{
Kinematic-mesh origin of the star test. The interval \(T=\{1,2\}\) sets
\(c_{13}=c_{14}=c_{15}=0\), while \(T=\{1,2,3\}\) sets
\(c_{14}=c_{15}=c_{24}=c_{25}=0\); the corresponding blocks are shaded. Only $X_{ij}$ within the shaded region appear in the respective zero conditions.}
    \label{fig:mesh}
\end{figure}

The origin of pollution is transparent in the kinematic mesh,
Fig.~\ref{fig:mesh}.  Each planar channel \(X_{ij}\) occupies a
mesh vertex, while each elementary diamond carries
\begin{equation}
 c_{ij}\equiv
 X_{ij}+X_{i+1,j+1}-X_{i,j+1}-X_{i+1,j}
 =-2p_i\!\cdot p_j .
 \label{eq:mesh-variable}
\end{equation}
An interval zero \(H_T\) sets \(c_{ij}=0\) throughout a rectangular block of
diamonds, while the channels \(X_{ij}\) themselves remain generic.  Summing
the elementary diamond relations over any rectangle within this block gives
\begin{equation}
 X_{ru}+X_{r'u'}=X_{ru'}+X_{r'u},
 \qquad r,r'\in T,\quad u,u'\notin T .
 \label{eq:mesh-rectangle}
\end{equation}

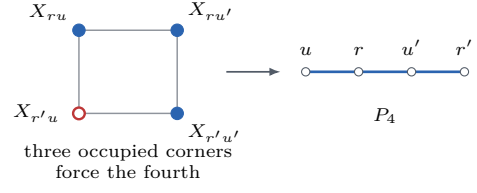
\begin{figure}[t]
\centering
\begin{tikzpicture}[
  font=\scriptsize,
  meshline/.style={draw=cutgray!65,line width=.55pt},
  chosen/.style={circle,fill=cutblue,draw=cutblue,inner sep=1.65pt},
  forced/.style={circle,draw=cutred,line width=.9pt,fill=white,
                 inner sep=1.55pt},
  graphline/.style={draw=cutblue,line width=1pt},
  graphpoint/.style={circle,draw=cutgray,fill=white,inner sep=1.1pt}
]
% Mesh rectangle
\coordinate (A) at (-2.65,.55);
\coordinate (B) at (-1.35,.55);
\coordinate (C) at (-2.65,-.55);
\coordinate (D) at (-1.35,-.55);
\draw[meshline] (A)--(B)--(D)--(C)--cycle;
\node[chosen] at (A) {};
\node[chosen] at (B) {};
\node[chosen] at (D) {};
\node[forced] at (C) {};
\node[above left=1pt]  at (A) {\(X_{ru}\)};
\node[above right=1pt] at (B) {\(X_{ru'}\)};
\node[left=4pt]        at (C) {\(X_{r'u}\)};
\node[below right=1pt] at (D) {\(X_{r'u'}\)};
\node[align=center] at (-2,-1.18)
  {three occupied corners\\force the fourth};

% Arrow
\draw[caseflow] (-.70,0)--(.02,0);

% Associated P4
\coordinate (g1) at (.35,0);
\coordinate (g2) at (1.05,0);
\coordinate (g3) at (1.75,0);
\coordinate (g4) at (2.45,0);
\draw[graphline] (g1)--(g2)--(g3)--(g4);
\foreach \q in {g1,g2,g3,g4}
  \node[graphpoint] at (\q) {};
\node[above=3pt] at (g1) {\(u\)};
\node[above=3pt] at (g2) {\(r\)};
\node[above=3pt] at (g3) {\(u'\)};
\node[above=3pt] at (g4) {\(r'\)};
\node at (1.40,-.58) {\(P_4\)};
\end{tikzpicture}
\caption{\label{fig:mesh-pollution}
The mesh origin of pollution.  Three selected corners of a rectangle (shown in blue) force
the fourth channel to vanish (red).  Interpreting each \(X_{ru}\) as a link between
\(r\) and \(u\), the same three corners form the path \(P_4\).}
\end{figure}

Taking \(\operatorname{Res}_M\) sets every pole in \(M\) to zero.  If three
corners of one of these rectangles are occupied by selected poles, or by the
identically vanishing polygon edges bordering \(T\),
Eq.~\eqref{eq:mesh-rectangle} forces the fourth channel to vanish as well, see Fig.~\ref{fig:mesh-pollution}.
A second pole product can then contribute to the same residue, which is
precisely the pollution described above.

This observation has a useful graphical form.  For an interval \(T\), draw
every selected chord crossing between \(T\) and its complement, together
with the two polygon edges at the ends of \(T\). In Fig.~\ref{fig:mesh}, these chords are precisely the $X_{ij}$ that lie in the shaded region of the particular zero.  Regard each retained
channel \(X_{ru}\) as a link between \(r\) and \(u\), and call the resulting
bipartite graph \(\Gamma_T(S)\), where \(S=\operatorname{supp}M\).  The three corners in
Fig.~\ref{fig:mesh-pollution} are exactly a three-link path \(P_4\).

A connected bipartite graph contains no \(P_4\) precisely when all its links
share one vertex,  it is a star.  Consequently, if every connected component
of \(\Gamma_T(S)\) is a star, the selected pole conditions are independent,
no additional pole is forced to vanish, and the coefficient remains
unpolluted.  We obtain the \emph{star test}
\begin{equation}
 \boxed{\begin{gathered}
 \left.\cB_n\right|_{H_T}=0,
 \\
 \Gamma_T(S)\text{ is a union of stars}
 \\
 \Longrightarrow\quad
 a_M=0 .
 \end{gathered}}
 \label{eq:star-test}
\end{equation}
We call such a \(T\) a \emph{detecting interval} for \(M\).

At five points, for example, \(M=\{13,24\}\) and \(T=\{4,5\}\) give
\[
 \Gamma_T(S)=\{34,24\}\sqcup\{51\},
\]
a union of two stars, so the zero eliminates the coefficient of
\(1/(X_{13}X_{24})\).  For the local support \(\{13,14\}\), the same interval
instead produces the path \(3-4-1-5\); the zero does not isolate this
coefficient, but relates it to a neighboring triangulation.  The complete
extraction argument, including repeated poles, is given in Appendix~1.

The locality problem is now graphical. For every nontriangulating support, we must find at least one interval whose cut graph is a union of stars.  We construct such an interval recursively below.

\section{Finding the zero recursively}
\label{sec:recursion}

The star test verifies a proposed zero; it remains to find one.  Starting
from a nontriangulating support \(S_n\), we remove a suitable vertex \(p\),
apply the result recursively to the resulting \((n-1)\)-gon, and lift the
child zero interval back to the original polygon,
\begin{equation}
 S_n\longrightarrow S_{n-1}
 \longrightarrow T_{n-1}\longrightarrow T_n .
\end{equation}
Once a detecting interval \(T_{n-1}\) has been found, there are only two
possible lifts: either
\begin{equation}
 T_n=T_{n-1},
 \qquad\text{or}\qquad
 T_n=T_{n-1}\cup\{p\}.
\end{equation}
Thus the only question is whether \(p\) must be added to the child zero
interval.  Remarkably, the answer depends only on the local chord
configuration around \(p\).  The counting result below guarantees that a
suitable \(p\) always exists, reducing the all-multiplicity construction to
four local cases.

The recursion rests on a simple but crucial counting fact.  If a free
vertex exists, meaning that no selected chord ends there, we choose it.
Otherwise, let \(L\) be the number of leaves, namely vertices touched by
exactly one selected chord.  Every other vertex is touched at least twice,
and therefore
\begin{equation}
 2|S_n|\geq L+2(n-L).
\end{equation}
Since \(|S_n|\leq n-3\), it follows that
\begin{equation}
 \boxed{\text{no free vertex}\quad\Longrightarrow\quad L\geq6.}
 \label{eq:six-leaves}
\end{equation}
The result is essential.  At full chord count, a nontriangulating
support contains a crossing pair.  Protecting its four endpoints leaves at
least two leaves available for removal, so the crossing survives in the
child.  Below full chord count, removing any leaf preserves the chord
deficit.  Thus the child is nontriangulating in either case.  This simple
counting result is the key to the recursion: it guarantees a removable
vertex and reduces the all-multiplicity proof to the four local
configurations below.

Let \(a,p,b\) be consecutive vertices and let \(X_{ab}\) be the
short chord above \(p\).  The chosen vertex is either free or a leaf, and
\(X_{ab}\) is either absent or present.  There are therefore only four cases,
shown in Fig.~\ref{fig:four-local-cases}.  In F0 the zero interval
\(\{a,p\}\) detects the term directly.  In F1 one deletes \(p\) and
\(X_{ab}\), then lifts a detecting zero of the child.  In L0 one either
finds a nearby detecting zero directly or deletes \(p\) and its unique
incident chord before recursing.  In L1 the short and the leaf chord are
removed, and the leaf chord is reattached through \(a\) or through \(b\),
producing two possible children.

The lifting lemma states that, in F1 and L0, every detecting zero interval of
the prescribed child has a allowed lift that detects the parent.  In L1, choose
a detecting interval from each of the the two child configurations.  Either interval
may fail to lift because restoring \(p\) creates a \(P_4\), but the two-slide
lemma guarantees that they cannot both fail, at least one has a allowed
detecting lift to the parent. Table~\ref{tab:lift-rules} exhausts the local lifting possibilities. 

Thus every branch either produces a detecting
zero directly or reduces the problem to lower multiplicity.    The complete lifting rules, the proof of the
two-slide lemma, worked examples, and higher-multiplicity checks are given
in Appendix~2. Appendix~2 also proves that triangulations evade every star test.

\begin{table*}[t]
\caption{\label{tab:lift-rules}Complete lifting rules.}
\centering
\footnotesize
\renewcommand{\arraystretch}{1.12}
\setlength{\tabcolsep}{5pt}
\setlength{\arrayrulewidth}{0.45pt}

\begin{tabular}{|c|c|c|c|}
\hline

\multicolumn{4}{|c|}{
\parbox[c]{0.93\textwidth}{\centering
Let \(T=T_{n-1}\) and \(T^+=T\cup\{p\}\).
In the last column, assume wlog \(a\in T\) and \(b\notin T\).
 \(\Gamma(S)\) denotes the cut graph.}
}
\\
\hline

\parbox[c]{0.04\textwidth}{\centering\textbf{Case}}
&
\parbox[c]{0.25\textwidth}{\centering\textbf{Condition/reduction}}
&
\parbox[c]{0.27\textwidth}{\centering
\textbf{\(a,b\) on the same side}}
&
\parbox[c]{0.34\textwidth}{\centering
\textbf{\(a,b\) on opposite sides}}
\\
\hline

F0
&
\parbox[c]{0.25\textwidth}{\centering
\(\displaystyle p\text{ free},\quad X_{ab}\notin S_n\)}
&
\multicolumn{2}{c|}{
\parbox[c]{0.63\textwidth}{\centering
\(\displaystyle T_n=\{a,p\}\)}
}
\\
\hline

F1
&
\parbox[c]{0.25\textwidth}{\centering
\(\displaystyle
\begin{gathered}
p\text{ free},\quad X_{ab}\in S_n,\\[-1pt]
S_{n-1}=S_n\setminus\{X_{ab}\}
\end{gathered}\)}
&
\parbox[c]{0.27\textwidth}{\centering
\(\displaystyle
\begin{gathered}
a,b\in T:\quad T_n=T^+,\\[-1pt]
a,b\notin T:\quad T_n=T
\end{gathered}\)}
&
\parbox[c]{0.34\textwidth}{\centering
\(\displaystyle
\begin{gathered}
\nexists\,X_{bv}\in\Gamma_T(S_{n-1}),\ v\neq a:
   \quad T_n=T,\\[-1pt]
\nexists\,X_{av}\in\Gamma_T(S_{n-1}),\ v\neq b:
   \quad T_n=T^+
\end{gathered}\)}
\\
\hline

\multirow{3}{*}{L0}
&
\parbox[c]{0.25\textwidth}{\centering
\(\displaystyle X_{au}\notin\Gamma(S_n)\)}
&
\multicolumn{2}{c|}{
\parbox[c]{0.63\textwidth}{\centering
\(\displaystyle T_n=\{a,p\}\)}
}
\\
\cline{2-4}

&
\parbox[c]{0.25\textwidth}{\centering
\(\displaystyle X_{bu}\notin\Gamma(S_n)\)}
&
\multicolumn{2}{c|}{
\parbox[c]{0.63\textwidth}{\centering
\(\displaystyle T_n=\{p,b\}\)}
}
\\
\cline{2-4}

&
\parbox[c]{0.25\textwidth}{\centering
\(\displaystyle
\begin{gathered}
X_{au},X_{bu}\in\Gamma(S_n),\\[-1pt]
S_{n-1}=S_n\setminus\{X_{pu}\}
\end{gathered}\)}
&
\parbox[c]{0.27\textwidth}{\centering
\(\displaystyle
\begin{gathered}
a,b\in T:\quad T_n=T^+,\\[-1pt]
a,b\notin T:\quad T_n=T
\end{gathered}\)}
&
\parbox[c]{0.34\textwidth}{\centering
\(\displaystyle
\begin{gathered}
u\in T:\quad T_n=T^+,\\[-1pt]
u\notin T:\quad T_n=T
\end{gathered}\)}
\\
\hline

L1
&
\parbox[c]{0.25\textwidth}{\centering
\(\displaystyle
\begin{gathered}
R=S_n\setminus\{X_{pu},X_{ab}\},\\[-1pt]
S_a=R\cup\{X_{au}\},\\[-1pt]
S_b=R\cup\{X_{bu}\}
\end{gathered}\)}
&
\parbox[c]{0.27\textwidth}{\centering
\(\displaystyle
\begin{gathered}
a,b\in T:\quad T_n=T^+,\\[-1pt]
a,b\notin T:\quad T_n=T
\end{gathered}\)}
&
\parbox[c]{0.34\textwidth}{\centering
\(\displaystyle
\begin{gathered}
u\in T:\quad \text{ first try }T^+\text{ then if it fails }T,\\[-1pt]
u\notin T:\quad \text{ first try } T\text{ then if it fails }T^+
\end{gathered}\)}
\\
\hline

\multicolumn{4}{|c|}{
\parbox[c]{0.93\textwidth}{\centering
Apply first to \(S_a\). If neither lift works, repeat with \(S_b\);
the two-slide lemma guarantees success.}
}
\\
\hline

\end{tabular}
\end{table*}

The construction is therefore directly executable. Given any excluded pole
product, we deterministically obtain a hidden-zero interval that isolates its coefficient.
A Wolfram Language implementation that constructs the child configurations,
tests their zero intervals, and displays every attempted lift and its
star-test certificate is included as ancillary material.  The lifting lemmas were also exhaustively checked through \(n=12\);
the corresponding case counts are given in Table~\ref{tab:recursion-checks}.

\section{Outlook}

This result closes an all-multiplicity question about emergent locality first
raised in \cite{GaugeLocality,HiddenZeros}.  More strongly than a uniqueness theorem,
it provides a direct method for extracting local graph structure from
regular kinematic zeros. 

For local ansatze, hidden zeros are equivalent to dual shuffle
factorization into \(D\)-subsets
\cite{CosmologicalWavefunctionZeros,ZhouShuffleFactorization}.  Without
assuming locality, our result suggests that this factorization is the local
limit of a more general structure in which zeros organize both local and
nonlocal pole products into related classes.

The immediate extensions are the NLSM and Yang--Mills theory, which are
organized by the same planar variables and possess related hidden zeros
\cite{UniversalZeros,NLSMRecursion,NLSMinTrphi,CirclesTriangles,
ScalarScaffoldedGluons,SurfaceYM}.  At loop level, the surface framework represents planar diagrams by curves on
punctured disks and develops all-loop curve and cut-equation formulations
\cite{BinaryGeometries,CausalDiamonds,AllLoopCounting,AllLoopMultiplicity,
TropicalColored,SurfaceYukawa,CutEquation,AllOrderSplits}.  The corresponding
surface integrands obey ``big-mountain'' zeros, which already give low-point
evidence for emergent loop-level locality \cite{BackusRodina}. These extensions are currently ongoing.

The theorem has a purely combinatorial reading that may be of
independent interest.  Triangulations of a polygon are usually characterized by
what they contain, $n-3$ pairwise noncrossing chords, or by indexing the vertices of the associahedron \cite{Stasheff}, equivalently
the facets of the type-$A$ cluster complex \cite{FominZelevinsky}.  Here they are
characterized by what they evade. Among configurations of at most $n-3$ chords,
triangulations are exactly those for which no interval cut graph is a union of
stars.  We are not aware of this
characterization in the combinatorics literature.

More broadly, the proof demonstrates the power of surfaceology. It turns a
question that is elementary to state, yet has resisted conventional on-shell
methods, into a finite combinatorial problem.  In doing so, it reveals
structures in scattering amplitudes that become visible only through surface
geometry and regular kinematic zeros.

\vskip 1cm

\noindent \emph{Acknowledgements}: The accompanying computer code was written by OpenAI Codex under the authors' direction. LR is supported by the National Natural Science Foundation of China General Program No. 12475070 and the Beijing Natural Science Foundation International Scientist Project No. IS24014. JV is supported by a Chinese Government Scholarship-TUNEM and by SPIC. 
\bibliographystyle{apsrev4-2}
\bibliography{locality_ref}

@article{Amplituhedron,
  author  = {Arkani-Hamed, Nima and Trnka, Jaroslav},
  title   = {The Amplituhedron},
  journal = {JHEP},
  volume  = {10},
  number  = {2014},
  pages   = {030},
  year    = {2014}
}

@article{ABHY,
  author  = {Arkani-Hamed, Nima and Bai, Yuntao and He, Song and Yan, Gongwang},
  title   = {Scattering Forms and the Positive Geometry of Kinematics, Color and the Worldsheet},
  journal = {JHEP},
  volume  = {05},
  number  = {2018},
  pages   = {096},
  year    = {2018}
}

@article{GaugeLocality,
  author  = {Arkani-Hamed, Nima and Rodina, Laurentiu and Trnka, Jaroslav},
  title   = {Locality and Unitarity of Scattering Amplitudes from Singularities and Gauge Invariance},
  journal = {Phys. Rev. Lett.},
  volume  = {120},
  pages   = {231602},
  year    = {2018}
}

@article{AdlerUniqueness,
  author  = {Rodina, Laurentiu},
  title   = {Uniqueness from Gauge Invariance and the Adler Zero},
  journal = {JHEP},
  volume  = {09},
  number  = {2019},
  pages   = {084},
  year    = {2019}
}

@article{SoftUniqueness,
  author  = {Rodina, Laurentiu},
  title   = {Scattering Amplitudes from Soft Theorems and Infrared Behavior},
  journal = {Phys. Rev. Lett.},
  volume  = {122},
  pages   = {071601},
  year    = {2019}
}

@article{RodinaUV,
  author  = {Rodina, Laurentiu},
  title   = {Hidden Zeros Are Equivalent to Enhanced Ultraviolet Scaling and Lead to Unique Amplitudes in Tr($\phi^3$) Theory},
  journal = {Phys. Rev. Lett.},
  volume  = {134},
  pages   = {031601},
  year    = {2025}
}

@article{HiddenZeros,
  author  = {Arkani-Hamed, Nima and Cao, Qu and Dong, Jin and Figueiredo, Carolina and He, Song},
  title   = {Hidden Zeros for Particle/String Amplitudes and the Unity of Colored Scalars, Pions and Gluons},
  journal = {JHEP},
  volume  = {10},
  number  = {2024},
  pages   = {231},
  year    = {2024}
}

@article{BackusRodina,
  author  = {Backus, Jeffrey V. and Rodina, Laurentiu},
  title   = {Emergence of Unitarity and Locality from Hidden Zeros at One-Loop Order},
  journal = {Phys. Rev. Lett.},
  volume  = {135},
  pages   = {131601},
  year    = {2025}
}

@article{UniversalZeros,
  author  = {Huang, Huajian and Yang, Yi-Jian and Zhou, Kang},
  title   = {Note on Hidden Zeros and Expansions of Tree-Level Amplitudes},
  journal = {Eur. Phys. J. C},
  volume  = {85},
  pages   = {685},
  year    = {2025}
}

@article{NLSMRecursion,
  author  = {Li, Xuan and Zhou, Kang},
  title   = {A New Recursion Relation for Tree-Level NLSM Amplitudes Based on Hidden Zeros},
  journal = {JHEP},
  volume  = {01},
  number  = {2026},
  pages   = {010},
  year    = {2026}
}

@article{Stasheff,
  author  = {Stasheff, James D.},
  title   = {Homotopy Associativity of H-Spaces. I and II},
  journal = {Trans. Am. Math. Soc.},
  volume  = {108},
  pages   = {275--292 and 293--333},
  year    = {1963}
}

@article{FominZelevinsky,
  author  = {Fomin, Sergey and Zelevinsky, Andrei},
  title   = {Y-Systems and Generalized Associahedra},
  journal = {Ann. Math.},
  volume  = {158},
  pages   = {977--1018},
  year    = {2003}
}

@article{BrownM0n,
  author  = {Brown, Francis C. S.},
  title   = {Multiple Zeta Values and Periods of Moduli Spaces $\mathcal{M}_{0,n}$},
  journal = {Ann. Sci. {\'E}c. Norm. Sup{\'e}r.},
  volume  = {42},
  pages   = {371--489},
  year    = {2009}
}

@misc{ZhouShuffleFactorization,
  author        = {Zhou, Kang},
  title         = {Universal Interpretation of Hidden Zero and
                   {$2$}-Split of Tree-Level Amplitudes Using Feynman
                   Diagrams, Part {I}: {$\operatorname{Tr}(\phi^3)$},
                   {NLSM} and {YM}},
  year          = {2026},
  eprint        = {2604.23680},
  archivePrefix = {arXiv},
  primaryClass  = {hep-th}
}

@article{RodinaBCFW,
  author  = {Rodina, Laurentiu},
  title   = {Uniqueness from Locality and BCFW Shifts},
  journal = {JHEP},
  volume  = {09},
  number  = {2019},
  pages   = {078},
  year    = {2019}
}

@article{CarrascoRodinaUV,
  author  = {Carrasco, John Joseph M. and Rodina, Laurentiu},
  title   = {UV Considerations on Scattering Amplitudes in a Web of Theories},
  journal = {Phys. Rev. D},
  volume  = {100},
  pages   = {125007},
  year    = {2019}
}

@article{NLSMRecursionOld,
  author  = {Kampf, Karol and Novotny, Jiri and Trnka, Jaroslav},
  title   = {Recursion Relations for Tree-Level Amplitudes in the $SU(N)$ Non-Linear Sigma Model},
  journal = {Phys. Rev. D},
  volume  = {87},
  pages   = {081701},
  year    = {2013}
}

@article{NLSMTree,
  author  = {Kampf, Karol and Novotny, Jiri and Trnka, Jaroslav},
  title   = {Tree-Level Amplitudes in the Nonlinear Sigma Model},
  journal = {JHEP},
  volume  = {05},
  number  = {2013},
  pages   = {032},
  year    = {2013}
}

@article{OnShellRecursionEFT,
  author  = {Cheung, Clifford and Kampf, Karol and Novotny, Jiri and Shen, Chia-Hsien and Trnka, Jaroslav},
  title   = {On-Shell Recursion Relations for Effective Field Theories},
  journal = {Phys. Rev. Lett.},
  volume  = {116},
  pages   = {041601},
  year    = {2016}
}

@article{EFTPeriodicTable,
  author  = {Cheung, Clifford and Kampf, Karol and Novotny, Jiri and Shen, Chia-Hsien and Trnka, Jaroslav},
  title   = {A Periodic Table of Effective Field Theories},
  journal = {JHEP},
  volume  = {02},
  number  = {2017},
  pages   = {020},
  year    = {2017}
}

@article{PositiveGeometries,
  author  = {Arkani-Hamed, Nima and Bai, Yuntao and Lam, Thomas},
  title   = {Positive Geometries and Canonical Forms},
  journal = {JHEP},
  volume  = {11},
  number  = {2017},
  pages   = {039},
  year    = {2017}
}

@article{BinaryGeometries,
  author  = {Arkani-Hamed, Nima and He, Song and Lam, Thomas and Thomas, Hugh},
  title   = {Binary Geometries, Generalized Particles and Strings, and Cluster Algebras},
  journal = {Phys. Rev. D},
  volume  = {107},
  pages   = {066015},
  year    = {2023}
}

@article{HiddenZerosDoubleCopy,
  author  = {Bartsch, Christian and Brown, T. V. and Kampf, Karol and Oktem, U. and Paranjape, S. and Trnka, Jaroslav},
  title   = {Hidden Amplitude Zeros from the Double-Copy Map},
  journal = {Phys. Rev. D},
  volume  = {111},
  pages   = {045019},
  year    = {2025}
}

@article{ScaffoldedGravityZeros,
  author  = {Li, Y. and Roest, D. and ter Veldhuis, T.},
  title   = {Hidden Zeros in Scaffolded General Relativity and Exceptional Field Theories},
  journal = {JHEP},
  volume  = {04},
  number  = {2025},
  pages   = {121},
  year    = {2025}
}

@article{ZhouDiagramZeros,
  author  = {Zhou, Kang},
  title   = {Understanding Zeros and Splittings of Ordered Tree Amplitudes via Feynman Diagrams},
  journal = {JHEP},
  volume  = {03},
  number  = {2025},
  pages   = {154},
  year    = {2025}
}

@article{FengBCFWZeros,
  author  = {Feng, Bo and Zhang, L. and Zhou, Kang},
  title   = {Hidden Zeros and 2-Split via BCFW Recursion Relation},
  journal = {JHEP},
  volume  = {08},
  number  = {2025},
  pages   = {205},
  year    = {2025}
}

@article{SmoothSplittingZeros,
  author  = {Jones, C. R. T. and Paranjape, S.},
  title   = {Smooth Splitting and Zeros from On-Shell Recursion},
  journal = {JHEP},
  volume  = {07},
  number  = {2025},
  pages   = {251},
  year    = {2025}
}

@article{CosmologicalHiddenZeros,
  author  = {De, S. and Paranjape, S. and Pokraka, A. and Spradlin, M. and Volovich, A.},
  title   = {Hidden Zeros of the Cosmological Wavefunction},
  journal = {JHEP},
  volume  = {07},
  number  = {2025},
  pages   = {174},
  year    = {2025}
}

@misc{CosmologicalWavefunctionZeros,
  author        = {Li, Yang and Rodina, Laurentiu},
  title         = {Cosmological Wavefunctions as Amplitudes: Dual Shuffle Factorization and Uniqueness from New Hidden Zeros},
  eprint        = {2604.01133},
  archivePrefix = {arXiv},
  primaryClass  = {hep-th},
  year          = {2026}
}

@misc{HiddenZerosMassive,
  author        = {Carrillo Gonz{\'a}lez, Mariana and Ward, F.},
  title         = {Hidden Zeros in Massive Theories},
  eprint        = {2601.16860},
  archivePrefix = {arXiv},
  primaryClass  = {hep-th},
  year          = {2026}
}

@article{HiddenZerosHigherDerivative,
  author  = {Zhou, Kang},
  title   = {Hidden Zeros for Higher-Derivative Yang--Mills and Gravity Amplitudes at Tree Level},
  journal = {JHEP},
  volume  = {02},
  number  = {2026},
  pages   = {039},
  year    = {2026}
}

@article{CausalDiamonds,
  author  = {Arkani-Hamed, Nima and He, Song and Salvatori, Giulio and Thomas, Hugh},
  title   = {Causal Diamonds, Cluster Polytopes and Scattering Amplitudes},
  journal = {JHEP},
  volume  = {11},
  number  = {2022},
  pages   = {049},
  year    = {2022}
}

@article{AllLoopCounting,
  author  = {Arkani-Hamed, Nima and Frost, Hadleigh and Salvatori, Giulio and Plamondon, Pierre-Guy and Thomas, Hugh},
  title   = {All Loop Scattering as a Counting Problem},
  journal = {JHEP},
  volume  = {08},
  number  = {2025},
  pages   = {194},
  year    = {2025}
}

@article{AllLoopMultiplicity,
  author  = {Arkani-Hamed, Nima and Frost, Hadleigh and Salvatori, Giulio and Plamondon, Pierre-Guy and Thomas, Hugh},
  title   = {All Loop Scattering for All Multiplicity},
  journal = {JHEP},
  volume  = {09},
  number  = {2025},
  pages   = {033},
  year    = {2025}
}

@article{TropicalColored,
  author  = {Arkani-Hamed, Nima and Figueiredo, Carolina and Frost, Hadleigh and Salvatori, Giulio},
  title   = {Tropical Amplitudes for Colored Lagrangians},
  journal = {JHEP},
  volume  = {05},
  number  = {2025},
  pages   = {051},
  year    = {2025}
}

@article{SurfaceYukawa,
  author  = {De, S. and Pokraka, A. and Skowronek, Marcos and Spradlin, M. and Volovich, A.},
  title   = {Surfaceology for Colored Yukawa Theory},
  journal = {JHEP},
  volume  = {09},
  number  = {2024},
  pages   = {160},
  year    = {2024}
}

@article{CutEquation,
  author  = {Arkani-Hamed, Nima and Frost, Hadleigh and Salvatori, Giulio},
  title   = {The Cut Equation},
  journal = {JHEP},
  volume  = {06},
  number  = {2026},
  pages   = {102},
  year    = {2026}
}

@article{NLSMinTrphi,
  author  = {Arkani-Hamed, Nima and Cao, Qu and Dong, Jin and Figueiredo, Carolina and He, Song},
  title   = {Nonlinear Sigma Model Amplitudes to All Loop Orders Are Contained in the Tr($\Phi^3$) Theory},
  journal = {Phys. Rev. D},
  volume  = {110},
  pages   = {065018},
  year    = {2024}
}

@article{CirclesTriangles,
  author  = {Arkani-Hamed, Nima and Figueiredo, Carolina},
  title   = {Circles and Triangles, the NLSM and Tr($\Phi^3$)},
  journal = {JHEP},
  volume  = {09},
  number  = {2025},
  pages   = {189},
  year    = {2025}
}

@article{ScalarScaffoldedGluons,
  author  = {Arkani-Hamed, Nima and Cao, Qu and Dong, Jin and Figueiredo, Carolina and He, Song},
  title   = {Scalar-Scaffolded Gluons and the Combinatorial Origins of Yang--Mills Theory},
  journal = {JHEP},
  volume  = {04},
  number  = {2025},
  pages   = {078},
  year    = {2025}
}

@article{SurfaceYM,
  author  = {Arkani-Hamed, Nima and Cao, Qu and Dong, Jin and Figueiredo, Carolina and He, Song},
  title   = {Surface Kinematics and the Canonical Yang--Mills All-Loop Integrand},
  journal = {Phys. Rev. Lett.},
  volume  = {134},
  pages   = {171601},
  year    = {2025}
}

@article{AllOrderSplits,
  author  = {Arkani-Hamed, Nima and Figueiredo, Carolina},
  title   = {All-Order Splits and Multi-Soft Limits for Particle and String Amplitudes},
  journal = {JHEP},
  volume  = {10},
  number  = {2025},
  pages   = {077},
  year    = {2025}
}

@article{CheungKampfNovotnyTrnka,
  author  = {Cheung, Clifford and Kampf, Karol and Novotny, Jiri and Trnka, Jaroslav},
  title   = {Effective Field Theories from Soft Limits of Scattering Amplitudes},
  journal = {Phys. Rev. Lett.},
  volume  = {114},
  number  = {22},
  pages   = {221602},
  year    = {2015}
}

\clearpage
\onecolumngrid
\appendix
\setcounter{section}{1}
\setcounter{equation}{0}
\renewcommand{\thesection}{\arabic{section}}
\renewcommand{\thesubsection}{\thesection.\arabic{subsection}}
\renewcommand{\theequation}{\arabic{section}.\arabic{equation}}

\section*{Appendix 1: Proof of the star test}
\label{app:star-test}

We prove the star test for the full ansatz \eqref{eq:ansatz}, including
repeated poles and terms with fewer than \(n-3\) denominator factors.  Fix a
zero interval
\begin{equation}
 T=\{w,w+1,\ldots,w+k\},\qquad v=w+k+1,
 \qquad
 2\leq |T|\leq n-3,
 \label{eq:app-zero-interval}
\end{equation}
and let \(T^{\rm c}\) denote its complement.  For a pole multiset \(M\), let
\(S=\operatorname{supp}M\) be the set of distinct chords appearing in \(M\).
A chord belonging to \(S\) will be called selected.  The cut graph
\(\Gamma_T(S)\) contains the selected chords crossing between \(T\) and
\(T^{\rm c}\), together with the two polygon edges at the ends of \(T\).
We prove
\begin{equation}
 \Gamma_T(S)\text{ is a union of stars},
 \qquad
 \left.\cB_n\right|_{H_T}=0
 \quad\Longrightarrow\quad
 a_M=0.
 \label{eq:app-star-test}
\end{equation}
We work in generic formal kinematics, without dimension-dependent Gram
relations.

\emph{Coordinates on the zero.---}
On \(H_T\), every channel crossing the cut can be written as
\begin{equation}
 X_{ru}=A_r+B_u,
 \qquad
 r\in T,\quad u\in T^{\rm c}.
 \label{eq:app-potential}
\end{equation}
The two polygon edges at the ends of \(T\) impose
\begin{equation}
 A_w+B_{w-1}=0,
 \qquad
 A_{w+k}+B_v=0.
 \label{eq:app-boundary}
\end{equation}
Explicitly, one may take
\begin{equation}
 A_r=X_{r,v},
 \qquad
 B_u=X_{w,u}-X_{w,v},
\end{equation}
which solves the zero relations \eqref{eq:zero}.  Equation~\eqref{eq:app-potential}
then gives
\begin{equation}
 X_{ru}+X_{r'u'}=X_{ru'}+X_{r'u},
 \label{eq:app-rectangle}
\end{equation}
which is the rectangle relation used in the main text.  Channels whose
endpoints lie on the same side of the cut remain independent.

\emph{Graph criterion.---}
Temporarily set to zero every link of \(\Gamma_T(S)\), including its two
boundary links.  Writing \(C_u=-B_u\), these conditions become
\(A_r=C_u\).  They are independent precisely when the cut graph is a forest.
Moreover, they force an additional channel \(X_{ru}\) to vanish precisely
when \(r\) and \(u\) lie in the same connected component.

It follows that no unselected channel is forced to vanish exactly when each
connected component contains every possible link between its two sets of
vertices, that is, when each component is complete bipartite.  A connected
complete bipartite graph is also a tree precisely when one side contains a
single vertex.  Such a graph is a star.  Therefore
\begin{equation}
 \boxed{
 \begin{gathered}
 \Gamma_T(S)\text{ is a union of stars}
 \\[2pt]
 \Longleftrightarrow
 \\[-2pt]
 \text{the selected pole conditions are independent and}\\
 \text{force no additional channel to vanish.}
 \end{gathered}}
 \label{eq:app-independent-flat}
\end{equation}
This is also the precise form of the mesh argument: three occupied corners
of a rectangle form a \(P_4\) and force the fourth corner to vanish.

\emph{Isolation of the coefficient.---}
Let \(C_\alpha\) be the connected components of \(\Gamma_T(S)\), including
isolated vertices, and choose distinct constants \(\lambda_\alpha\).  Introduce
a large parameter \(\Lambda\) and set
\begin{equation}
 A_r=\lambda_\alpha\Lambda+a_r,
 \qquad
 B_u=-\lambda_\alpha\Lambda+b_u,
 \qquad r,u\in C_\alpha .
 \label{eq:app-large-shift}
\end{equation}
This deformation remains entirely inside \(H_T\).  Because each nontrivial
component is a star, its finite parameters may be chosen so that all selected
channels crossing the cut take arbitrary independent values.  We also assign
independent variables to all channels lying entirely on one side of the cut.

Every selected channel remains finite.  By
Eq.~\eqref{eq:app-independent-flat}, an unselected channel crossing the cut
joins two different components and therefore grows as
\begin{equation}
 X_{ru}
 =
 (\lambda_\alpha-\lambda_\beta)\Lambda+a_r+b_u
 =
 O(\Lambda),
 \qquad \alpha\neq\beta .
 \label{eq:app-growing-channel}
\end{equation}
Since the numerators in \eqref{eq:ansatz} are constant, every term containing
such a channel vanishes as \(\Lambda\to\infty\).

The restricted zero condition consequently becomes a Laurent identity in
the independent finite channels.  Distinct pole multisets give distinct
Laurent monomials, so the coefficient of each monomial must vanish
separately.  In particular, the monomial with precisely the pole
multiplicities of \(M\) has coefficient \(a_M\), and hence \(a_M=0\).

Repeated poles are distinguished by their Laurent exponents, while terms
with fewer factors have different zero exponents.  Thus the same argument
covers the full ansatz.  The bound \(|M|\leq n-3\) is not required for the
star test itself; it enters only in Appendix~2, where it guarantees the
existence of a detecting interval.

\setcounter{section}{2}
\setcounter{equation}{0}

\section*{Appendix 2: Recursive construction}
\label{app:recursive-proof}

\tikzset{
  casechildedge/.style={line width=.95pt,draw=cutorange,dashed},
  atlaspoint/.style={circle,draw=cutgray,fill=white,line width=.55pt,
    minimum size=4.0mm,inner sep=0pt,font=\scriptsize},
  atlasfree/.style={circle,draw=cutorange,fill=cutorange!14,line width=.8pt,
    minimum size=4.2mm,inner sep=0pt,font=\scriptsize},
  atlasleaf/.style={circle,draw=cutred,fill=cutred!10,line width=.8pt,
    minimum size=4.2mm,inner sep=0pt,font=\scriptsize},
  atlascenter/.style={circle,draw=cutgreen!85!black,fill=cutgreen!18,
    line width=.8pt,minimum size=4.0mm,inner sep=0pt,font=\scriptsize},
  atlasstage/.style={font=\small\bfseries,text=cutgray,align=center},
  atlasnote/.style={font=\scriptsize,align=center,text=cutgray},
  atlasboundary/.style={draw=cutorange,line width=1.0pt},
  atlasbad/.style={draw=cutred,line width=1.35pt},
  atlasabsent/.style={draw=cutgray!55,line width=.75pt,densely dotted},
  atlasgoodtext/.style={font=\scriptsize\bfseries,text=cutgreen!75!black},
  atlasbadtext/.style={font=\scriptsize\bfseries,text=cutred}
}

We now prove that every nontriangulating support within the chord bound has a
detecting interval.  For the cut-graph problem, \(T\) and \(T^{\rm c}\) define
the same partition; we always choose the representative satisfying
\(2\leq |T|\leq n-3\).  A zero interval is detecting when every connected
component of its cut graph is a star.  We prove
\begin{equation}
 |S|\leq n-3,
 \qquad
 S\text{ is not a triangulation}
 \quad\Longrightarrow\quad
 \exists\,T:\ \Gamma_T(S)\text{ is a union of stars}.
 \label{app:star-cut-theorem}
\end{equation}
Pole multiplicities play no role here because the cut graph depends only on
the support \(S\).

\subsection{Induction and allowed lifts}

There are two induction branches.  In the \emph{deficit branch},
\(|S|\leq n-4\).  In the \emph{protected-crossing branch},
\(|S|\leq n-3\) and \(S\) contains a chosen crossing pair whose four
endpoints are protected throughout the reduction.  Every support in
Eq.~\eqref{app:star-cut-theorem} belongs to at least one branch. If
\(|S|=n-3\), a nontriangulating support must contain a crossing pair.

Let \(a,p,b\) be consecutive vertices of the current polygon.  After deleting
\(p\), the vertices \(a\) and \(b\) become adjacent; we call \(ab\) the new
child boundary edge.  A parent interval \(T_n\) lifts a child interval
\(T_{n-1}\) if deleting \(p\) from \(T_n\) gives \(T_{n-1}\).  There are at
most two possibilities:
\begin{equation}
 T_n=T_{n-1},
 \qquad\text{or}\qquad
 T_n=T_{n-1}\cup\{p\}.
 \label{app:possible-lifts}
\end{equation}
Only choices that remain cyclic intervals are allowed.

If \(a,b\in T_{n-1}\), the unique lift is
\(T_n=T_{n-1}\cup\{p\}\).  If \(a,b\notin T_{n-1}\), the unique lift is
\(T_n=T_{n-1}\).  If exactly one of \(a,b\) belongs to \(T_{n-1}\), both
choices are possible.

The induction begins at five points.  At full chord count, every
nontriangulating two-chord support is a crossing pair.  Its fifth vertex is
free, and the short chord above that vertex is absent, so the F0 construction
below applies.  Supports with at most one chord are treated in the same way
after choosing a free vertex whose short chord is absent.

We next show that a suitable vertex always exists.  A vertex is
\emph{free} if no selected chord ends there and is a \emph{leaf} if exactly
one selected chord ends there.  If there is no free vertex and \(L\) vertices
are leaves, then
\begin{equation}
 2|S|=\sum_v\deg_S(v)
 \geq L+2(n-L).
\end{equation}
Since \(|S|\leq n-3\),
\begin{equation}
 L\geq6.
 \label{app:leaf-count}
\end{equation}
In the deficit branch, deleting any leaf and its unique chord leaves at most
\(n-5\) chords, so the child still has a chord deficit.  In the
protected-crossing branch, choose a leaf away from the four protected
endpoints; the six-leaf bound guarantees at least two choices.  Deleting this
leaf preserves the protected crossing.

The short chord \(X_{ab}\) is never part of the protected pair.  Indeed, any
chord crossing \(X_{ab}\) must end at \(p\), because \(X_{ab}\) cuts off the
triangle \(apb\).  Since \(p\) is not a protected endpoint, \(X_{ab}\) cannot
be protected.  Thus every child constructed below satisfies the same
inductive condition as its parent.

The four cases are determined by whether \(p\) is free or a leaf and whether
the short chord \(X_{ab}\) is absent or present.

\subsection{F0: Free vertex, short absent}

Suppose \(p\) is free and \(X_{ab}\notin S\).  Then
\begin{equation}
 T_n=\{a,p\}
 \label{app:F0-cut}
\end{equation}
is detecting.  Every selected chord crossing this interval ends at \(a\),
since none ends at \(p\).  These chords and the polygon edge immediately
preceding \(a\) therefore form a star centered at \(a\).  The boundary edge
\(pb\) is a separate one-edge star, and the absence of \(X_{ab}\) prevents
the two components from joining.

This is the direct five-point construction illustrated in the main text, so
no additional figure is needed here.

\subsection{F1: Free vertex, short present}

Suppose \(p\) is free and \(X_{ab}\in S\).  Delete \(p\) and \(X_{ab}\),
giving
\begin{equation}
 S_{n-1}=S\setminus\{X_{ab}\}.
 \label{app:F1-child}
\end{equation}
Let \(T_{n-1}\) be any detecting interval of the child.

If \(a,b\in T_{n-1}\), choose
\(T_n=T_{n-1}\cup\{p\}\).  If \(a,b\notin T_{n-1}\), choose
\(T_n=T_{n-1}\).  In either case \(a,p,b\) all lie on the same side of the
parent cut, so the restored chord \(X_{ab}\) does not appear in the parent
cut graph, which is unchanged.

Now suppose \(a\in T_{n-1}\) and \(b\notin T_{n-1}\).  Then \(ab\) is a
boundary link of the child cut graph.  Because that graph is a union of
stars, other links crossing the cut can end at \(a\) or at \(b\), but not at
both, otherwise they and \(ab\) would form a \(P_4\).

If the other links end at \(a\), choose \(T_n=T_{n-1}\).  The child boundary
link \(ab\) is then replaced by the parent boundary link \(ap\) and the
restored chord \(X_{ab}\), both incident on \(a\).  If the other links end
at \(b\), choose \(T_n=T_{n-1}\cup\{p\}\); the corresponding two links are
now \(pb\) and \(X_{ab}\), both incident on \(b\).  If neither endpoint has
another link, either lift works.  Hence every detecting interval of the
child has an allowed detecting lift.

\paragraph*{Example.}
For \(S_6=\{13,15,24\}\), choose the free vertex \(p=6\).  Deleting \(6\)
and the short chord \(15\) gives \(S_5=\{13,24\}\).  The detecting interval
\(T_5=\{4,5\}\) crosses the new child edge \(51\).  Choosing
\(T_6=T_5\) replaces the child link \(51\) by the two links \(56\) and
\(15\), both centered at \(5\).

\FloatBarrier
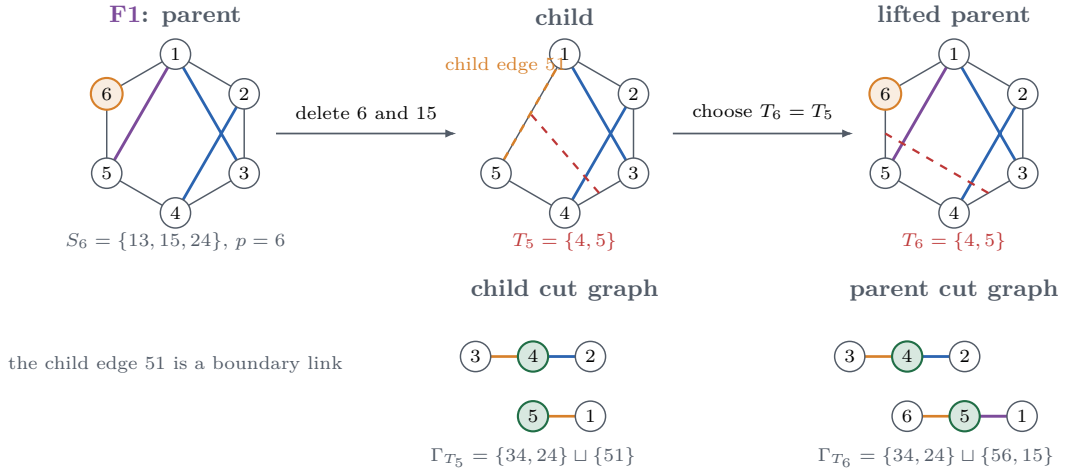
\begin{figure}[!ht]
\centering
\begin{tikzpicture}[x=1cm,y=1cm,font=\scriptsize]
  \begin{scope}[shift={(-5.15,.85)}]
    \node[atlasstage] at (0,1.55) {\textcolor{cutpurple}{F1}: parent};
    \foreach \i/\ang in {1/90,2/30,3/-30,4/-90,5/-150,6/150}
      \coordinate (FB\i) at (\ang:1.05);
    \draw[casepoly] (FB1)--(FB2)--(FB3)--(FB4)--(FB5)--(FB6)--cycle;
    \draw[casesupp] (FB1)--(FB3) (FB2)--(FB4);
    \draw[caseshort] (FB1)--(FB5);
    \foreach \i in {1,...,5} \node[atlaspoint] at (FB\i) {\i};
    \node[atlasfree] at (FB6) {6};
    \node[atlasnote] at (0,-1.42) {$S_6=\{13,15,24\}$, $p=6$};
  \end{scope}
  \draw[caseflow] (-3.82,.85)--(-1.43,.85)
    node[midway,above=2pt,align=center,font=\scriptsize]
      {delete $6$ and $15$};

  \begin{scope}[shift={(0,.85)}]
    \node[atlasstage] at (0,1.55) {child};
    \foreach \i/\ang in {1/90,2/30,3/-30,4/-90,5/-150}
      \coordinate (FBC\i) at (\ang:1.05);
    \draw[casepoly] (FBC1)--(FBC2)--(FBC3)--(FBC4)--(FBC5)--cycle;
    \draw[casesupp] (FBC1)--(FBC3) (FBC2)--(FBC4);
    \draw[casechildedge] (FBC5)--(FBC1);
    \draw[casecut] ($(FBC3)!.5!(FBC4)$)--($(FBC5)!.5!(FBC1)$);
    \foreach \i in {1,...,5} \node[atlaspoint] at (FBC\i) {\i};
    \node[atlasnote,text=cutorange] at (-.78,.90) {child edge $51$};
    \node[atlasnote,text=cutred] at (0,-1.42) {$T_5=\{4,5\}$};
  \end{scope}
  \draw[caseflow] (1.43,.85)--(3.82,.85)
    node[midway,above=2pt,align=center,font=\scriptsize]
      {choose $T_6=T_5$};

  \begin{scope}[shift={(5.15,.85)}]
    \node[atlasstage] at (0,1.55) {lifted parent};
    \foreach \i/\ang in {1/90,2/30,3/-30,4/-90,5/-150,6/150}
      \coordinate (FBP\i) at (\ang:1.05);
    \draw[casepoly] (FBP1)--(FBP2)--(FBP3)--(FBP4)--(FBP5)--(FBP6)--cycle;
    \draw[casesupp] (FBP1)--(FBP3) (FBP2)--(FBP4);
    \draw[caseshort] (FBP1)--(FBP5);
    \draw[casecut] ($(FBP3)!.5!(FBP4)$)--($(FBP5)!.5!(FBP6)$);
    \foreach \i in {1,...,5} \node[atlaspoint] at (FBP\i) {\i};
    \node[atlasfree] at (FBP6) {6};
    \node[atlasnote,text=cutred] at (0,-1.42) {$T_6=\{4,5\}$};
  \end{scope}

  \begin{scope}[shift={(0,-2.25)}]
    \node[atlasstage] at (0,1.03) {child cut graph};
    \coordinate (FBg3) at (-1.18,.15);
    \coordinate (FBg4) at (-.42,.15);
    \coordinate (FBg2) at (.34,.15);
    \coordinate (FBg5) at (-.42,-.64);
    \coordinate (FBg1) at (.34,-.64);
    \draw[atlasboundary] (FBg3)--(FBg4);
    \draw[casesupp] (FBg4)--(FBg2);
    \draw[atlasboundary] (FBg5)--(FBg1);
    \node[atlaspoint] at (FBg3) {3};
    \node[atlascenter] at (FBg4) {4};
    \node[atlaspoint] at (FBg2) {2};
    \node[atlascenter] at (FBg5) {5};
    \node[atlaspoint] at (FBg1) {1};
    \node[atlasnote] at (-.42,-1.18)
      {$\Gamma_{T_5}=\{34,24\}\sqcup\{51\}$};
  \end{scope}
  \begin{scope}[shift={(5.15,-2.25)}]
    \node[atlasstage] at (0,1.03) {parent cut graph};
    \coordinate (FBh3) at (-1.38,.15);
    \coordinate (FBh4) at (-.62,.15);
    \coordinate (FBh2) at (.14,.15);
    \coordinate (FBh6) at (-.62,-.64);
    \coordinate (FBh5) at (.14,-.64);
    \coordinate (FBh1) at (.90,-.64);
    \draw[atlasboundary] (FBh3)--(FBh4);
    \draw[casesupp] (FBh4)--(FBh2);
    \draw[atlasboundary] (FBh6)--(FBh5);
    \draw[caseshort] (FBh5)--(FBh1);
    \node[atlaspoint] at (FBh3) {3};
    \node[atlascenter] at (FBh4) {4};
    \node[atlaspoint] at (FBh2) {2};
    \node[atlaspoint] at (FBh6) {6};
    \node[atlascenter] at (FBh5) {5};
    \node[atlaspoint] at (FBh1) {1};
    \node[atlasnote] at (-.24,-1.18)
      {$\Gamma_{T_6}=\{34,24\}\sqcup\{56,15\}$};
  \end{scope}
  \node[atlasnote] at (-5.15,-2.20)
    {the child edge $51$ is a boundary link};
\end{tikzpicture}
\caption{F1 when the new child edge $51$ is a boundary link of the child cut graph.  Choosing $T_6=T_5=\{4,5\}$ leaves both $6$ and $1$ outside the interval.  The one-edge star $51$ becomes the two-arm star $6-5-1$.}
\label{fig:app-atlas-F1-boundary}
\end{figure}
\FloatBarrier

\subsection{L0: Leaf, short absent}

Let \(X_{pu}\) be the unique selected chord ending at \(p\), and suppose
\(X_{ab}\notin S\).

If \(X_{au}\) is neither a selected chord nor a polygon edge, then
\(T_n=\{a,p\}\) is detecting.  The links ending at \(a\) form one star, while
\(X_{pu}\) and the boundary edge \(pb\) form another star centered at \(p\).
The two components cannot join through \(u\), because \(X_{au}\) is absent,
or through \(b\), because \(X_{ab}\) is absent.  If instead \(X_{bu}\) is
neither a selected chord nor a polygon edge, the reflected interval
\(T_n=\{p,b\}\) is detecting.

It remains to consider the case in which both \(X_{au}\) and \(X_{bu}\) are
either selected chords or polygon edges.  Delete \(p\) and \(X_{pu}\), giving
\begin{equation}
 S_{n-1}=S\setminus\{X_{pu}\},
 \label{app:L0-child}
\end{equation}
and let \(T_{n-1}\) be any detecting interval of the child.

First suppose \(a\) and \(b\) lie on the same side of the child cut.  Use the
unique lift in Eq.~\eqref{app:possible-lifts}.  If \(p\) and \(u\) lie on
the same side of the parent cut, \(X_{pu}\) does not appear in its cut graph.
Otherwise both \(X_{au}\) and \(X_{bu}\) cross the child cut.  Since the child
graph is a union of stars, their component must be centered at \(u\).
Restoring \(X_{pu}\) then adds one more arm to that same star.

Now suppose \(a\in T_{n-1}\) and \(b\notin T_{n-1}\).  If
\(u\notin T_{n-1}\), choose \(T_n=T_{n-1}\).  Then \(p\) and \(u\) are both
outside \(T_n\), so \(X_{pu}\) does not cross the parent cut.  In the child,
the links \(X_{au}\) and \(ab\) force their component to be centered at
\(a\); replacing the boundary link \(ab\) by the new boundary link \(ap\)
preserves that star.  If \(u\in T_{n-1}\), choose
\(T_n=T_{n-1}\cup\{p\}\).  Now \(X_{pu}\) again stays on one side, while
\(X_{bu}\) and \(ab\) force the child star to be centered at \(b\); replacing
\(ab\) by \(pb\) preserves it.  Thus every detecting child interval again
has an allowed detecting lift.

\paragraph*{Direct example.}
For \(S_6=\{14,25,36\}\), choose the leaf \(p=6\), whose unique chord is
\(36\).  The short chord \(15\) and the link \(35\) are absent, so
\(T_6=\{5,6\}\) detects the parent directly:
\[
 \Gamma_{T_6}(S_6)
 =
 \{45,25\}\sqcup\{61,36\}.
\]

\FloatBarrier
\begin{figure}[!ht]
\centering
\begin{tikzpicture}[x=1cm,y=1cm,font=\scriptsize]
  \begin{scope}[shift={(-2.75,.25)}]
    \node[atlasstage] at (0,1.55)
      {\textcolor{cutpurple}{L0}: parent and direct cut};
    \foreach \i/\ang in {1/90,2/30,3/-30,4/-90,5/-150,6/150}
      \coordinate (LD\i) at (\ang:1.05);
    \draw[casepoly] (LD1)--(LD2)--(LD3)--(LD4)--(LD5)--(LD6)--cycle;
    \draw[casesupp] (LD1)--(LD4) (LD2)--(LD5);
    \draw[caseleaf] (LD3)--(LD6);
    \draw[atlasabsent,draw=cutpurple] (LD1)--(LD5);
    \draw[atlasabsent,draw=cutorange] (LD1)--(LD3) (LD3)--(LD5);
    \draw[casecut] ($(LD4)!.5!(LD5)$)--($(LD6)!.5!(LD1)$);
    \foreach \i in {1,...,5} \node[atlaspoint] at (LD\i) {\i};
    \node[atlasleaf] at (LD6) {6};
    \node[atlasnote,text=cutred] at (0,-1.42) {$T_6=\{5,6\}$};
    \node[atlasnote,text=cutorange] at (0,.12) {$13,35\notin S_6$};
  \end{scope}
  \draw[caseflow] (-1.12,.25)--(.16,.25)
    node[midway,above=2pt,font=\scriptsize] {star test};
  \begin{scope}[shift={(2.55,.25)}]
    \node[atlasstage] at (0,1.55) {parent cut graph};
    \coordinate (LDg4) at (-1.46,.30);
    \coordinate (LDg5) at (-.67,.30);
    \coordinate (LDg2) at (.12,.30);
    \coordinate (LDg1) at (-.67,-.62);
    \coordinate (LDg6) at (.12,-.62);
    \coordinate (LDg3) at (.91,-.62);
    \draw[atlasboundary] (LDg4)--(LDg5);
    \draw[casesupp] (LDg5)--(LDg2);
    \draw[atlasboundary] (LDg1)--(LDg6);
    \draw[caseleaf] (LDg6)--(LDg3);
    \node[atlaspoint] at (LDg4) {4};
    \node[atlascenter] at (LDg5) {5};
    \node[atlaspoint] at (LDg2) {2};
    \node[atlaspoint] at (LDg1) {1};
    \node[atlascenter] at (LDg6) {6};
    \node[atlaspoint] at (LDg3) {3};
    \node[atlasnote] at (-.27,-1.28)
      {$\Gamma_{T_6}=\{45,25\}\sqcup\{61,36\}$};
    \node[atlasgoodtext] at (-.27,-1.66) {two stars};
  \end{scope}
\end{tikzpicture}
\caption{The direct L0 move.  The missing link $35$ makes the displayed
two-vertex interval detecting; the reflected construction uses the missing
link $13$.}
\label{fig:app-atlas-L0-direct}
\end{figure}
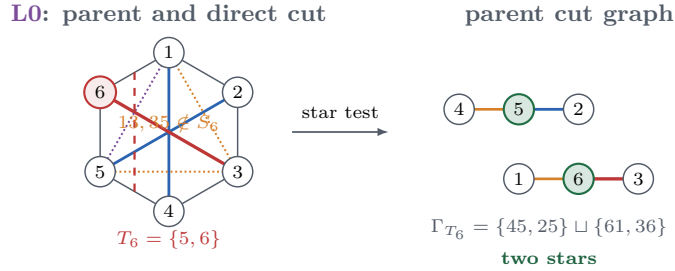
\FloatBarrier

\paragraph*{Recursive example.}
For \(S_6=\{13,24,36\}\), choose \(p=1\), with unique chord \(13\).
Deleting \(1\) and \(13\) gives \(S_5=\{24,36\}\).  The interval
\(T_5=\{5,6\}\) crosses the new child edge \(62\).  Since \(u=3\) lies
outside \(T_5\), choose \(T_6=T_5\), so \(p=1\) also lies outside the
parent interval.  The restored chord \(13\) is then absent from the cut
graph, while the child boundary link \(62\) is replaced by \(61\).

\FloatBarrier
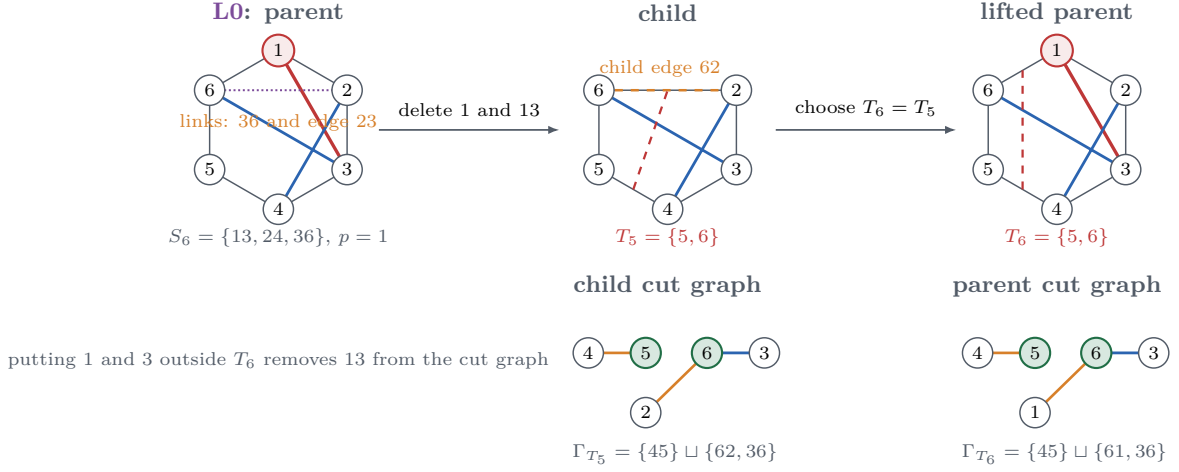
\begin{figure}[!ht]
\centering
\begin{tikzpicture}[x=1cm,y=1cm,font=\scriptsize]
  \begin{scope}[shift={(-5.15,.85)}]
    \node[atlasstage] at (0,1.55) {\textcolor{cutpurple}{L0}: parent};
    \foreach \i/\ang in {1/90,2/30,3/-30,4/-90,5/-150,6/150}
      \coordinate (LB\i) at (\ang:1.05);
    \draw[casepoly] (LB1)--(LB2)--(LB3)--(LB4)--(LB5)--(LB6)--cycle;
    \draw[caseleaf] (LB1)--(LB3);
    \draw[casesupp] (LB2)--(LB4) (LB3)--(LB6);
    \draw[atlasabsent,draw=cutpurple] (LB2)--(LB6);
    \node[atlasleaf] at (LB1) {1};
    \foreach \i in {2,...,6} \node[atlaspoint] at (LB\i) {\i};
    \node[atlasnote] at (0,-1.42) {$S_6=\{13,24,36\}$, $p=1$};
    \node[atlasnote,text=cutorange] at (0,.10) {links: $36$ and edge $23$};
  \end{scope}
  \draw[caseflow] (-3.82,.85)--(-1.43,.85)
    node[midway,above=2pt,font=\scriptsize] {delete $1$ and $13$};

  \begin{scope}[shift={(0,.85)}]
    \node[atlasstage] at (0,1.55) {child};
    \foreach \i/\ang in {2/30,3/-30,4/-90,5/-150,6/150}
      \coordinate (LBC\i) at (\ang:1.05);
    \draw[casepoly] (LBC2)--(LBC3)--(LBC4)--(LBC5)--(LBC6)--cycle;
    \draw[casesupp] (LBC2)--(LBC4) (LBC3)--(LBC6);
    \draw[casechildedge] (LBC6)--(LBC2);
    \draw[casecut] ($(LBC4)!.5!(LBC5)$)--($(LBC6)!.5!(LBC2)$);
    \foreach \i in {2,...,6} \node[atlaspoint] at (LBC\i) {\i};
    \node[atlasnote,text=cutorange] at (-.10,.82) {child edge $62$};
    \node[atlasnote,text=cutred] at (0,-1.42) {$T_5=\{5,6\}$};
  \end{scope}
  \draw[caseflow] (1.43,.85)--(3.82,.85)
    node[midway,above=2pt,align=center,font=\scriptsize]
      {choose $T_6=T_5$};

  \begin{scope}[shift={(5.15,.85)}]
    \node[atlasstage] at (0,1.55) {lifted parent};
    \foreach \i/\ang in {1/90,2/30,3/-30,4/-90,5/-150,6/150}
      \coordinate (LBP\i) at (\ang:1.05);
    \draw[casepoly] (LBP1)--(LBP2)--(LBP3)--(LBP4)--(LBP5)--(LBP6)--cycle;
    \draw[caseleaf] (LBP1)--(LBP3);
    \draw[casesupp] (LBP2)--(LBP4) (LBP3)--(LBP6);
    \draw[casecut] ($(LBP4)!.5!(LBP5)$)--($(LBP6)!.5!(LBP1)$);
    \node[atlasleaf] at (LBP1) {1};
    \foreach \i in {2,...,6} \node[atlaspoint] at (LBP\i) {\i};
    \node[atlasnote,text=cutred] at (0,-1.42) {$T_6=\{5,6\}$};
  \end{scope}

  \begin{scope}[shift={(0,-2.25)}]
    \node[atlasstage] at (0,1.03) {child cut graph};
    \coordinate (LBg4) at (-1.05,.15);
    \coordinate (LBg5) at (-.29,.15);
    \coordinate (LBg2) at (-.29,-.64);
    \coordinate (LBg6) at (.52,.15);
    \coordinate (LBg3) at (1.28,.15);
    \draw[atlasboundary] (LBg4)--(LBg5);
    \draw[atlasboundary] (LBg2)--(LBg6);
    \draw[casesupp] (LBg6)--(LBg3);
    \node[atlaspoint] at (LBg4) {4};
    \node[atlascenter] at (LBg5) {5};
    \node[atlaspoint] at (LBg2) {2};
    \node[atlascenter] at (LBg6) {6};
    \node[atlaspoint] at (LBg3) {3};
    \node[atlasnote] at (.10,-1.18)
      {$\Gamma_{T_5}=\{45\}\sqcup\{62,36\}$};
  \end{scope}
  \begin{scope}[shift={(5.15,-2.25)}]
    \node[atlasstage] at (0,1.03) {parent cut graph};
    \coordinate (LBh4) at (-1.05,.15);
    \coordinate (LBh5) at (-.29,.15);
    \coordinate (LBh1) at (-.29,-.64);
    \coordinate (LBh6) at (.52,.15);
    \coordinate (LBh3) at (1.28,.15);
    \draw[atlasboundary] (LBh4)--(LBh5);
    \draw[atlasboundary] (LBh1)--(LBh6);
    \draw[casesupp] (LBh6)--(LBh3);
    \node[atlaspoint] at (LBh4) {4};
    \node[atlascenter] at (LBh5) {5};
    \node[atlaspoint] at (LBh1) {1};
    \node[atlascenter] at (LBh6) {6};
    \node[atlaspoint] at (LBh3) {3};
    \node[atlasnote] at (.10,-1.18)
      {$\Gamma_{T_6}=\{45\}\sqcup\{61,36\}$};
  \end{scope}
  \node[atlasnote] at (-5.15,-2.20)
    {putting $1$ and $3$ outside $T_6$ removes $13$ from the cut graph};
\end{tikzpicture}
\caption{L0 when the new child edge $62$ is a boundary link of the child cut graph.  Choosing $T_6=T_5=\{5,6\}$ places both $p=1$ and $u=3$ outside the interval, so $13$ does not cross the cut.  The boundary link $62$ is replaced by $61$.}
\label{fig:app-atlas-L0-boundary}
\end{figure}
\FloatBarrier

\subsection{L1: Leaf, short present}

Finally suppose \(X_{pu},X_{ab}\in S\).  Delete \(p\), \(X_{pu}\), and
\(X_{ab}\), and write
\begin{equation}
 R=S\setminus\{X_{pu},X_{ab}\}.
 \label{app:L1-remainder}
\end{equation}
Form two child supports
\begin{equation}
 S_a=R\cup\{X_{au}\},
 \qquad
 S_b=R\cup\{X_{bu}\}.
 \label{app:L1-children}
\end{equation}
We call these the \(a\)-child and \(b\)-child.  If the added link is a polygon
edge, it is omitted from the support; if it already belongs to \(R\), it is
not duplicated.  Both children therefore contain at most \(n-4\) selected
chords and preserve the relevant induction condition.

Choose arbitrary detecting intervals \(T_a\) and \(T_b\) for the two
children.  To reconstruct the parent, restore \(p\), \(X_{pu}\), and
\(X_{ab}\).  A link \(X_{zu}\), \(z=a,b\), that was introduced only in the
child is removed; if it already belonged to \(R\), it remains selected in
the parent.

A detecting interval of one child need not lift.  For example, suppose the
\(a\)-child contains an additional selected link \(X_{ay}\), so its cut graph
contains the star \(y-a-u\).  If \(X_{au}\in R\), this link remains in the
parent.  Restoring \(X_{pu}\) then creates the path \(y-a-u-p\), which is a
\(P_4\).  The role of the second child is to avoid this obstruction.

\medskip
\noindent\textbf{Two-slide lemma.}
\emph{For any detecting intervals \(T_a\) and \(T_b\) of the two children,
at least one has an allowed lift for which the parent cut graph is a union
of stars.}

\smallskip
\noindent\emph{Proof.}
For \(z\in\{a,b\}\), let \(\bar z\) denote the other vertex in
\(\{a,b\}\).  Call \(T_z\) obstructed for the \(z\)-child if none of its
allowed parent lifts is detecting.

First suppose \(a\) and \(b\) lie on the same side of \(T_z\), so the lift is
unique.  If \(u\) lies on that same side, neither restored chord crosses the
parent cut and the lift succeeds.  If \(u\) lies on the opposite side and
\(X_{zu}\) was introduced only in the child, the lift simply replaces the
child link \(X_{zu}\) by \(X_{pu}\).  If the child star is centered at \(u\),
one arm is exchanged for another; if it is centered at \(z\), removing
\(X_{zu}\) leaves \(X_{pu}\) as a separate one-edge star.  The lift again
succeeds.

An obstruction with \(a\) and \(b\) on the same side is therefore possible
only when \(X_{zu}\) remains in the parent, either because
\begin{equation}
 \begin{aligned}
 \mathrm{(I_s)}&\quad X_{zu}\in R,
 \\
 \mathrm{(I_e)}&\quad X_{zu}\text{ is a polygon edge}.
 \end{aligned}
 \label{app:internal-obstructions}
\end{equation}
In either case, the child component containing \(X_{zu}\) must be centered
at \(z\) and must contain a second link \(X_{z y_z}\), where \(y_z\) denotes
its other endpoint.  Restoring \(X_{pu}\) then creates
\(y_z-z-u-p\).  In case \(\mathrm{(I_e)}\), the second link \(X_{z y_z}\)
must be selected; if it were also a boundary link, one side of the zero
interval would contain only \(z\), which is not allowed.

Now suppose \(ab\) is a boundary link of the child cut graph.  If \(u\) lies
on the same side as \(\bar z\), choose the lift that places \(p\) on that
side.  Then \(X_{pu}\) does not cross the parent cut.  The restored short
\(X_{ab}\) and the new parent boundary link both end at \(z\).  The child
link \(X_{zu}\), together with the boundary link \(ab\), already forces the
child component to be centered at \(z\), so this replacement preserves a
star.

An obstructed interval must therefore place \(u\) on the same side as \(z\).
Testing the other lift shows that failure is possible only if
\begin{equation}
 \mathrm{(B)}\qquad
 ab\text{ belongs to a star centered at }z
 \text{ containing another selected link }X_{z y_z}.
 \label{app:boundary-obstruction}
\end{equation}
The parent then contains the path \(y_z-z-\bar z-p\).  Thus every obstructed
\(z\)-interval separates \(\bar z\) from \(u\).

Assume, for contradiction, that both \(T_a\) and \(T_b\) are obstructed.
The pair \((\mathrm{I_e},\mathrm{I_e})\) is impossible because \(u\) cannot
be a polygon neighbor of both consecutive vertices \(a\) and \(b\) for
\(n\geq5\).  The pairs
\((\mathrm{I_e},\mathrm{B})\) and
\((\mathrm{B},\mathrm{I_e})\) would leave one side of one interval with only
one vertex and are therefore also impossible.

Every remaining pairing involving an internal obstruction creates a \(P_4\)
in one of the two child cut graphs:
\begin{center}
\begin{tabular}{c c c}
\hline
type of \(T_a\)&type of \(T_b\)&forbidden child path\\
\hline
\(\mathrm{I_s}\)&\(\mathrm{I_s}\) or \(\mathrm{I_e}\)
  &\(a-u-b-y_b\subset\Gamma_{T_b}(S_b)\)\\
\(\mathrm{I_e}\)&\(\mathrm{I_s}\)
  &\(b-u-a-y_a\subset\Gamma_{T_a}(S_a)\)\\
\(\mathrm{I_s}\)&\(\mathrm{B}\)
  &\(u-a-b-y_b\subset\Gamma_{T_b}(S_b)\)\\
\(\mathrm{B}\)&\(\mathrm{I_s}\)
  &\(u-b-a-y_a\subset\Gamma_{T_a}(S_a)\)\\
\hline
\end{tabular}
\end{center}
For instance, in the first row the obstruction for \(T_a\) implies
\(X_{au}\in R\).  Since an obstructed \(b\)-interval separates \(a\) from
\(u\), this chord crosses the \(b\)-cut.  The obstruction for \(T_b\)
supplies \(X_{ub}\) and \(X_{b y_b}\), producing the displayed \(P_4\).
The other rows follow identically or by reflection.

It remains to exclude \((\mathrm{B},\mathrm{B})\).  For an interval whose
cut graph contains the boundary link \(ab\), let \(B_z\) denote the side of
\(T_z\) containing \(b\).  Such sides are cyclic intervals beginning at
\(b\), and hence \(B_a\) and \(B_b\) are nested.  The obstruction for
\(T_a\) gives \(u\notin B_a\), whereas the obstruction for \(T_b\) gives
\(u\in B_b\); therefore \(B_a\subsetneq B_b\).  It also gives
\(y_a\in B_a\), while \(a\notin B_b\), so \(X_{a y_a}\) crosses the
\(b\)-cut.  Together with \(X_{ab}\) and \(X_{b y_b}\), it forms
\[
 y_a-a-b-y_b\subset\Gamma_{T_b}(S_b),
\]
contradicting the assumption that \(T_b\) is detecting.

Thus \(T_a\) and \(T_b\) cannot both be obstructed.  At least one has an
allowed detecting lift to the parent.  \(\square\)

\paragraph*{Example.}
Let
\[
 S_7=\{14,16,27,35\},
 \qquad
 p=7,\quad a=6,\quad b=1,\quad u=2.
\]
After deleting \(7,27,16\), the \(b\)-child is
\(S_b=\{14,35\}\), since \(12\) is a polygon edge.  Its detecting interval
\(T_b=\{1,2\}\) has two allowed parent lifts.  The lift
\(T_7=\{1,2\}\) contains the \(P_4\), \(4-1-7-2\), while the lift
\(T_7=\{7,1,2\}\) contains the \(P_4\), \(7-6-1-4\); thus neither lift is
detecting.  The \(a\)-child is
\(S_a=\{14,26,35\}\).  Its interval \(T_a=\{2,3\}\) lifts with
\(T_7=T_a\), replacing the arm \(26\) by \(27\):
\[
 \Gamma_{T_a}(S_a)
 =\{12,26\}\sqcup\{34,35\},
 \qquad
 \Gamma_{T_7}(S_7)
 =\{12,27\}\sqcup\{34,35\}.
\]

\FloatBarrier
\begin{figure}[!ht]
\centering
\begin{tikzpicture}[x=1cm,y=1cm,font=\scriptsize]
  % ------------------------------ bad row, parent
  \begin{scope}[shift={(-5.15,3.55)}]
    \node[atlasstage] at (0,1.55) {\textcolor{cutpurple}{L1}: parent};
    \foreach \i/\ang in {1/90,2/38.571,3/-12.857,4/-64.286,5/-115.714,6/-167.143,7/141.429}
      \coordinate (LXB\i) at (\ang:1.05);
    \draw[casepoly] (LXB1)--(LXB2)--(LXB3)--(LXB4)--(LXB5)--(LXB6)--(LXB7)--cycle;
    \draw[casesupp] (LXB1)--(LXB4) (LXB3)--(LXB5);
    \draw[caseshort] (LXB1)--(LXB6);
    \draw[caseleaf] (LXB2)--(LXB7);
    \foreach \i in {1,...,6} \node[atlaspoint] at (LXB\i) {\i};
    \node[atlasleaf] at (LXB7) {7};
    \node[atlasnote] at (0,-1.42) {$S_7=\{14,16,27,35\}$};
  \end{scope}
  \draw[caseflow] (-3.82,3.55)--(-1.43,3.55)
    node[midway,above=2pt,align=center,font=\scriptsize]
      {$b$-child: $27\to12$\\[-2pt](polygon edge, omit)};

  % bad child
  \begin{scope}[shift={(0,3.55)}]
    \node[atlasstage] at (0,1.55) {$b$-child};
    \foreach \i/\ang in {1/90,2/38.571,3/-12.857,4/-64.286,5/-115.714,6/-167.143}
      \coordinate (LXBC\i) at (\ang:1.05);
    \draw[casepoly] (LXBC1)--(LXBC2)--(LXBC3)--(LXBC4)--(LXBC5)--(LXBC6)--cycle;
    \draw[casechildedge] (LXBC6)--(LXBC1);
    \draw[casesupp] (LXBC1)--(LXBC4) (LXBC3)--(LXBC5);
    \draw[casecut] ($(LXBC6)!.5!(LXBC1)$)--($(LXBC2)!.5!(LXBC3)$);
    \foreach \i in {1,...,6} \node[atlaspoint] at (LXBC\i) {\i};
    \node[atlasnote,text=cutred] at (0,-1.42) {$T_6^{(b)}=\{1,2\}$};
  \end{scope}
  \draw[caseflow] (1.43,3.55)--(3.82,3.55)
    node[midway,above=2pt,font=\scriptsize] {try $T_7=\{1,2\}$};

  % bad parent
  \begin{scope}[shift={(5.15,3.55)}]
    \node[atlasstage,text=cutred] at (0,1.55) {failed parent lift};
    \foreach \i/\ang in {1/90,2/38.571,3/-12.857,4/-64.286,5/-115.714,6/-167.143,7/141.429}
      \coordinate (LXBP\i) at (\ang:1.05);
    \draw[casepoly] (LXBP1)--(LXBP2)--(LXBP3)--(LXBP4)--(LXBP5)--(LXBP6)--(LXBP7)--cycle;
    \draw[casesupp] (LXBP1)--(LXBP4) (LXBP3)--(LXBP5);
    \draw[caseshort] (LXBP1)--(LXBP6);
    \draw[caseleaf] (LXBP2)--(LXBP7);
    \draw[casecut] ($(LXBP7)!.5!(LXBP1)$)--($(LXBP2)!.5!(LXBP3)$);
    \foreach \i in {1,...,6} \node[atlaspoint] at (LXBP\i) {\i};
    \node[atlasleaf] at (LXBP7) {7};
    \node[atlasnote,text=cutred] at (0,-1.42) {$T_7^{(b)}=\{1,2\}$};
  \end{scope}

  % child bad graph
  \begin{scope}[shift={(0,.48)}]
    \node[atlasstage] at (0,1.02) {child cut graph};
    \coordinate (LXBg6) at (-1.18,.15);
    \coordinate (LXBg1) at (-.42,.15);
    \coordinate (LXBg4) at (.34,.15);
    \coordinate (LXBg2) at (-.42,-.64);
    \coordinate (LXBg3) at (.34,-.64);
    \draw[atlasboundary] (LXBg6)--(LXBg1);
    \draw[casesupp] (LXBg1)--(LXBg4);
    \draw[atlasboundary] (LXBg2)--(LXBg3);
    \node[atlaspoint] at (LXBg6) {6};
    \node[atlascenter] at (LXBg1) {1};
    \node[atlaspoint] at (LXBg4) {4};
    \node[atlascenter] at (LXBg2) {2};
    \node[atlaspoint] at (LXBg3) {3};
    \node[atlasnote] at (-.42,-1.18)
      {$\Gamma_{T_6^{(b)}}=\{61,14\}\sqcup\{23\}$};
  \end{scope}
  % bad parent graph
  \begin{scope}[shift={(5.15,.48)}]
    \node[atlasstage,text=cutred] at (0,1.02) {parent cut graph};
    \coordinate (LXBh4) at (-1.65,.15);
    \coordinate (LXBh1) at (-.83,.15);
    \coordinate (LXBh6) at (-.83,-.65);
    \coordinate (LXBh7) at (0,.15);
    \coordinate (LXBh2) at (.83,.15);
    \coordinate (LXBh3) at (1.65,.15);
    \draw[casesupp] (LXBh4)--(LXBh1);
    \draw[caseshort] (LXBh1)--(LXBh6);
    \draw[atlasboundary] (LXBh1)--(LXBh7);
    \draw[caseleaf] (LXBh7)--(LXBh2);
    \draw[atlasboundary] (LXBh2)--(LXBh3);
    \node[atlaspoint] at (LXBh4) {4};
    \node[atlaspoint] at (LXBh1) {1};
    \node[atlaspoint] at (LXBh6) {6};
    \node[atlaspoint] at (LXBh7) {7};
    \node[atlaspoint] at (LXBh2) {2};
    \node[atlaspoint] at (LXBh3) {3};
    \draw[atlasbad] (LXBh4)--(LXBh1)--(LXBh7)--(LXBh2);
    \node[atlasbadtext] at (0,-1.16) {$P_4=4-1-7-2$};
  \end{scope}
  \node[atlasnote] at (-5.15,.50)
    {remove $7,27,16$};

  % ------------------------------ good row, parent
  \begin{scope}[shift={(-5.15,-3.25)}]
    \node[atlasstage] at (0,1.55) {\textcolor{cutpurple}{L1}: same parent};
    \foreach \i/\ang in {1/90,2/38.571,3/-12.857,4/-64.286,5/-115.714,6/-167.143,7/141.429}
      \coordinate (LXA\i) at (\ang:1.05);
    \draw[casepoly] (LXA1)--(LXA2)--(LXA3)--(LXA4)--(LXA5)--(LXA6)--(LXA7)--cycle;
    \draw[casesupp] (LXA1)--(LXA4) (LXA3)--(LXA5);
    \draw[caseshort] (LXA1)--(LXA6);
    \draw[caseleaf] (LXA2)--(LXA7);
    \foreach \i in {1,...,6} \node[atlaspoint] at (LXA\i) {\i};
    \node[atlasleaf] at (LXA7) {7};
    \node[atlasnote] at (0,-1.42) {$a=6$, $p=7$, $b=1$, $u=2$};
  \end{scope}
  \draw[caseflow] (-3.82,-3.25)--(-1.43,-3.25)
    node[midway,above=2pt,font=\scriptsize] {$a$-child: $27\to26$};

  % good child
  \begin{scope}[shift={(0,-3.25)}]
    \node[atlasstage] at (0,1.55) {$a$-child};
    \foreach \i/\ang in {1/90,2/38.571,3/-12.857,4/-64.286,5/-115.714,6/-167.143}
      \coordinate (LXAC\i) at (\ang:1.05);
    \draw[casepoly] (LXAC1)--(LXAC2)--(LXAC3)--(LXAC4)--(LXAC5)--(LXAC6)--cycle;
    \draw[casechildedge] (LXAC6)--(LXAC1);
    \draw[casesupp] (LXAC1)--(LXAC4) (LXAC2)--(LXAC6) (LXAC3)--(LXAC5);
    \draw[casecut] ($(LXAC1)!.5!(LXAC2)$)--($(LXAC3)!.5!(LXAC4)$);
    \foreach \i in {1,...,6} \node[atlaspoint] at (LXAC\i) {\i};
    \node[atlasnote,text=cutred] at (0,-1.42) {$T_6^{(a)}=\{2,3\}$};
  \end{scope}
  \draw[caseflow] (1.43,-3.25)--(3.82,-3.25)
    node[midway,above=2pt,font=\scriptsize] {lift $T_7=T_6^{(a)}$};

  % good parent
  \begin{scope}[shift={(5.15,-3.25)}]
    \node[atlasstage,text=cutgreen!75!black] at (0,1.55) {successful parent lift};
    \foreach \i/\ang in {1/90,2/38.571,3/-12.857,4/-64.286,5/-115.714,6/-167.143,7/141.429}
      \coordinate (LXAP\i) at (\ang:1.05);
    \draw[casepoly] (LXAP1)--(LXAP2)--(LXAP3)--(LXAP4)--(LXAP5)--(LXAP6)--(LXAP7)--cycle;
    \draw[casesupp] (LXAP1)--(LXAP4) (LXAP3)--(LXAP5);
    \draw[caseshort] (LXAP1)--(LXAP6);
    \draw[caseleaf] (LXAP2)--(LXAP7);
    \draw[casecut] ($(LXAP1)!.5!(LXAP2)$)--($(LXAP3)!.5!(LXAP4)$);
    \foreach \i in {1,...,6} \node[atlaspoint] at (LXAP\i) {\i};
    \node[atlasleaf] at (LXAP7) {7};
    \node[atlasnote,text=cutred] at (0,-1.42) {$T_7=\{2,3\}$};
  \end{scope}

  % good child graph
  \begin{scope}[shift={(0,-6.33)}]
    \node[atlasstage] at (0,1.02) {child cut graph};
    \coordinate (LXAg1) at (-1.22,.15);
    \coordinate (LXAg2) at (-.44,.15);
    \coordinate (LXAg6) at (-.44,-.64);
    \coordinate (LXAg4) at (.36,.15);
    \coordinate (LXAg3) at (1.14,.15);
    \coordinate (LXAg5) at (1.14,-.64);
    \draw[atlasboundary] (LXAg1)--(LXAg2);
    \draw[casesupp] (LXAg2)--(LXAg6);
    \draw[atlasboundary] (LXAg4)--(LXAg3);
    \draw[casesupp] (LXAg3)--(LXAg5);
    \node[atlaspoint] at (LXAg1) {1};
    \node[atlascenter] at (LXAg2) {2};
    \node[atlaspoint] at (LXAg6) {6};
    \node[atlaspoint] at (LXAg4) {4};
    \node[atlascenter] at (LXAg3) {3};
    \node[atlaspoint] at (LXAg5) {5};
    \node[atlasnote] at (-.02,-1.18)
      {$\Gamma_{T_6^{(a)}}=\{12,26\}\sqcup\{34,35\}$};
  \end{scope}
  % good parent graph
  \begin{scope}[shift={(5.15,-6.33)}]
    \node[atlasstage,text=cutgreen!75!black] at (0,1.02) {parent cut graph};
    \coordinate (LXAh1) at (-1.22,.15);
    \coordinate (LXAh2) at (-.44,.15);
    \coordinate (LXAh7) at (-.44,-.64);
    \coordinate (LXAh4) at (.36,.15);
    \coordinate (LXAh3) at (1.14,.15);
    \coordinate (LXAh5) at (1.14,-.64);
    \draw[atlasboundary] (LXAh1)--(LXAh2);
    \draw[caseleaf] (LXAh2)--(LXAh7);
    \draw[atlasboundary] (LXAh4)--(LXAh3);
    \draw[casesupp] (LXAh3)--(LXAh5);
    \node[atlaspoint] at (LXAh1) {1};
    \node[atlascenter] at (LXAh2) {2};
    \node[atlaspoint] at (LXAh7) {7};
    \node[atlaspoint] at (LXAh4) {4};
    \node[atlascenter] at (LXAh3) {3};
    \node[atlaspoint] at (LXAh5) {5};
    \node[atlasnote] at (-.02,-1.18)
      {$\Gamma_{T_7}=\{12,27\}\sqcup\{34,35\}$};
    \node[atlasgoodtext] at (-.02,-1.58) {two stars};
  \end{scope}
  \node[atlasnote] at (-5.15,-6.30)
    {the alternative child replaces $26$ by $27$};
\end{tikzpicture}
\caption{The two children in L1.  The upper row displays the obstructed
lift $T_7=\{1,2\}$ from the $b$-child; the other allowed lift,
$T_7=\{7,1,2\}$, is also obstructed, by $7-6-1-4$.  The $a$-child instead
supplies a detecting lift. It replaces the arm $26$ by $27$, so the parent
cut graph remains a union of stars.}
\label{fig:app-atlas-L1}
\end{figure}
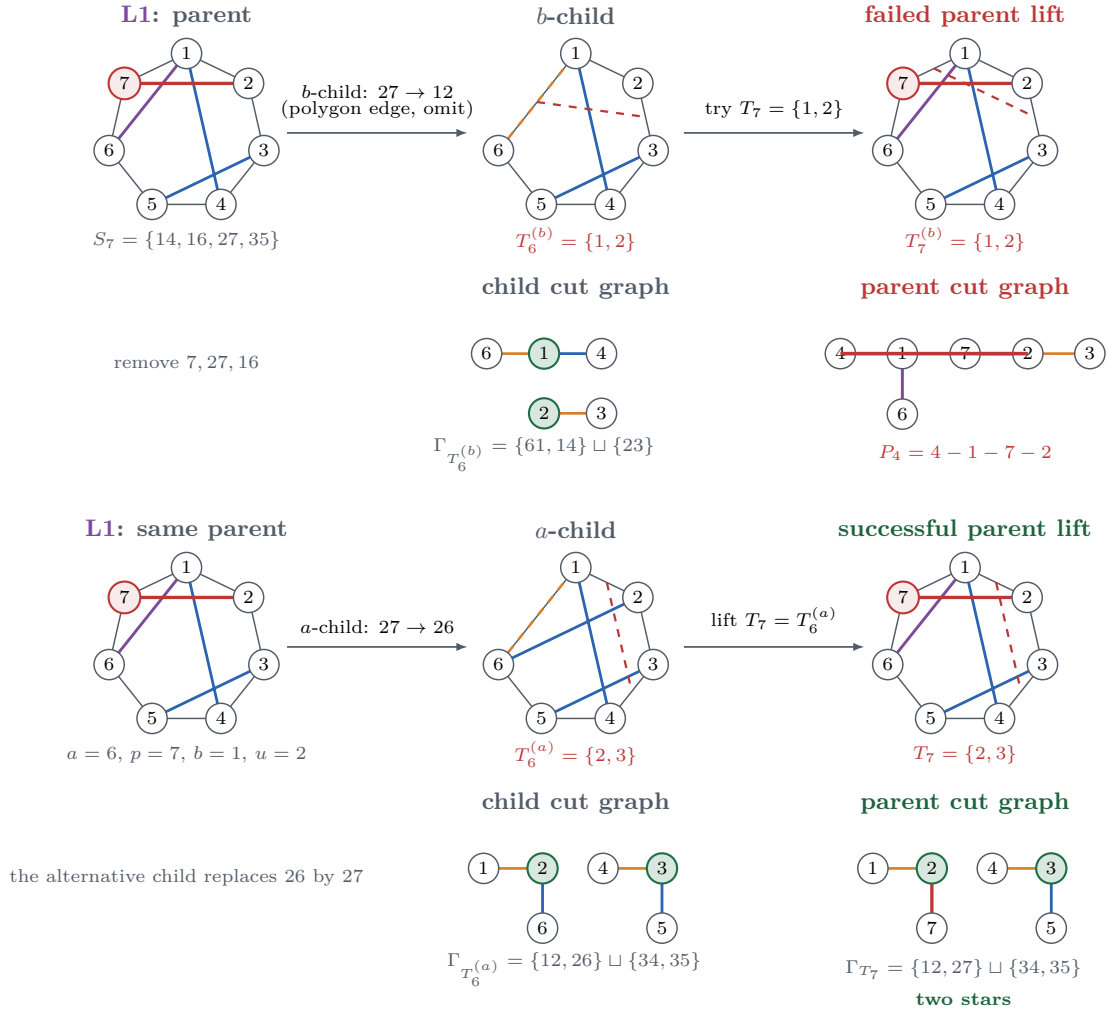
\FloatBarrier

\subsection{Completion}

The four cases are exhaustive because the chosen vertex is free or a leaf
and its short chord is absent or present.  F0 returns a detecting interval
directly.  In F1, every detecting interval of the prescribed child has an
allowed detecting lift.  In L0, either one of the two-vertex intervals works
directly or every detecting interval of the prescribed child lifts.  In L1,
the two-slide lemma guarantees that detecting intervals from the two
children cannot both be obstructed.  Every recursive step lowers the number
of vertices, so the construction reaches the five-point base case and proves
Eq.~\eqref{app:star-cut-theorem}.

The proof is constructive.  Fix an ordering of vertices, crossing pairs,
zero intervals, and allowed lifts.  At each stage choose the first free
vertex; if none exists, choose the first admissible leaf.  Apply the
appropriate rule above and recurse.  In L1, obtain one detecting interval
from each child and test their allowed lifts in the fixed order.  The
two-slide lemma guarantees that this finite procedure returns a detecting
zero interval.

\subsection{Why triangulations evade every test}

Let \(S\) triangulate the polygon and let \(T\) be any allowed zero interval.
Color the vertices in \(T\) and \(T^{\rm c}\) by \(0\) and \(1\), respectively,
and extend the coloring linearly across every triangle.  The level set
\(f=\tfrac12\) meets the polygon boundary at exactly the two boundary links
of the cut.  Inside each triangle it is a line segment joining two sides.
Because the dual graph of a polygon triangulation is a tree, these segments
cannot form a closed component.  They therefore form one arc joining the two
boundary links.

Consecutive links crossed by this arc belong to a common triangle and hence
share a vertex.  The two boundary links consequently lie in one connected
component of \(\Gamma_T(S)\).  They are disjoint because both sides of the
zero interval contain at least two vertices, so that component cannot be a
star.  No zero interval detects a triangulation.

Together with Eq.~\eqref{app:star-cut-theorem}, this proves that, among
supports containing at most \(n-3\) chords, triangulations are precisely
those that evade every interval star test.

\subsection{Explicit checks}

We counted exactly through twelve points all nonlocal pole multisets and all
admissible choices of the removed vertex.  A rooted recursion case is a pair
\((M,p)\), where \(p\) is a free vertex or, if none exists, a leaf whose
deletion preserves the relevant induction condition.  At full support size,
this means that at least one crossing pair survives the deletion.

The recursion depends only on \(S=\operatorname{supp}M\).  Nevertheless,
different multiplicities are counted as different terms.  A support of size
\(s\) has \(\binom{n-3}{s}\) positive multiplicity assignments of total
degree at most \(n-3\).

We also tested every detecting child interval and every allowed lift, finding
no counterexample to the lifting lemmas.  Individual L1 lifts can be
obstructed, but no simultaneous two-child obstruction was found.  OpenAI
Codex was used to assist with implementing and auditing these explicit
computer checks.

\begin{table}[h]
\centering
\caption{\label{tab:recursion-checks}
Exact numbers of nonlocal pole multisets \(M\) and admissible recursion choices \((M,p)\).  Repeated poles and terms with fewer than \(n-3\) factors are included.  A pole multiset may contribute more than once because several vertices \(p\) may be admissible.}
\begin{ruledtabular}
\begin{tabular}{r r r r r r}
\(n\) & \(N_M\) & F0 & F1 & L0 & L1 \\ \hline
5  & \(16\)              & \(30\)              & \(10\)              & \(0\)             & \(0\) \\
6  & \(206\)             & \(336\)             & \(138\)             & \(18\)            & \(0\) \\
7  & \(3{,}018\)         & \(5{,}005\)         & \(1{,}904\)         & \(308\)           & \(84\) \\
8  & \(52{,}998\)        & \(93{,}024\)        & \(30{,}672\)        & \(6{,}360\)       & \(1{,}872\) \\
9  & \(1{,}107{,}139\)   & \(2{,}072{,}070\)   & \(590{,}832\)       & \(142{,}614\)     & \(37{,}728\) \\
10 & \(26{,}976{,}898\)  & \(53{,}796{,}160\)  & \(13{,}444{,}750\)  & \(3{,}535{,}490\) & \(827{,}290\) \\
11 & \(752{,}533{,}288\) & \(1{,}595{,}093{,}643\)
                         & \(354{,}449{,}524\) & \(96{,}783{,}192\)
                         & \(20{,}231{,}948\) \\
12 & \(23{,}667{,}673{,}019\)
                         & \(53{,}179{,}362{,}600\)
                         & \(10{,}635{,}814{,}176\)
                         & \(2{,}912{,}891{,}436\)
                         & \(550{,}246{,}848\) \\
\end{tabular}
\end{ruledtabular}
\end{table}
\FloatBarrier

\end{document}